\documentclass[lettersize,journal]{IEEEtran}
\usepackage{amsmath,amsfonts,amssymb}
\usepackage{algpseudocode}
\usepackage{algorithm}
\usepackage{array}
\usepackage[caption=false,font=normalsize,labelfont=sf,textfont=sf]{subfig}
\usepackage{textcomp}
\usepackage{stfloats}
\usepackage{url}
\usepackage{verbatim}
\usepackage{cite}
\usepackage{optidef}
\usepackage{graphicx}
\usepackage{caption}
\usepackage{subfig}
\usepackage[T1]{fontenc}
\usepackage{multirow}
\usepackage{amsmath,amsthm}

\begin{document}

\title{Smart Energy Management with Optimized Prosumerism for Achieving Dynamic Net-Zero Balance in Electrified Road Transport Networks}

\author{Ferheen Ayaz, Maziar Nekovee
\thanks{F. Ayaz and M. Nekovee are with the 6G Lab, School of Engineering and Informatics, University of
        	Sussex, Brighton, BN1 9RH, U.K. e-mail: (f.ayaz@sussex.ac.uk, m.nekovee@sussex.ac.uk).}
\thanks{The research leading to this publication is funded by the UKRI/EPSRC Network Plus “A Green Connected and Prosperous	Britain”.}}



\maketitle

\begin{abstract}
The increasing number of Electric Vehicles (EVs) have led to rising energy demands which aggregates the burden on grid supply. A few solutions have been proposed to achieve demand and supply balance, for example, using storage systems for storing surplus energy from EVs or scheduling supply from the gird according to varying demand at different times. However, these solutions are costly and their applicability is limited to specific regions and times. This paper proposes a smart energy management solution for a massively electrified road transport network. It comprises of energy supplies from grid, charging stations, distributed renewable sources and EVs connected by 5G-enabled aggregators. We propose EVs as prosumers, which are energy consumers but also supply back their surplus energy via bidirectional Vehicle-to-Grid (V2G) technology. We use machine learning models to forecast hourly energy output from renewable sources, surplus supply from EVs and their demands. Two types of renewable sources i.e., wind and photovoltaic systems and three types of EVs, i.e., car, bus and lorry with different specifications and battery capacities are considered. A grid cost minimization solution is proposed using Mixed Integer Linear Programming which dynamically alters supply according to demand and energy provision from EVs. The proposed solution also considers penalty charge for CO$_2$ emissions during energy generation. The upper bounds of surplus supply and demand of EVs are theoretically derived. An incentive distribution mechanism is also presented to reward EVs offering their surplus supply and to discourage them to become selfish which is analyzed using Prisoner's dilemma game theory. Additionally, the paper presents an optimum number of charging stations on a road considering the incentives of EVs and their maximum contribution in supplying energy. Simulation results show that the proposed solution can effectively meet the demand requirements with increasing number of EVs, even if the supply from grid is limited, and can averagely reduce 38.21\% of grid load. The optimization with prosumer EVs results in 5.3\% of average cost reduction compared with optimization without prosumers. Also, the penalty charge for CO$_2$ emissions results in over 50\% cost reduction by using renewable resources in the proposed solution as compared to fossil fuels. The communication and computation complexity of the proposed solution is shown to be reduced with the exploitation of 5G-enabled aggregator.
\end{abstract}

\begin{IEEEkeywords}
EV, prosumer, energy management, optimization, MILP, emissions, V2G, prisoner's dilemma.
\end{IEEEkeywords}

\section{Introduction}
\IEEEPARstart{E}{lectric} Vehicles (EVs) are becoming an essential part of road transportation due to their positive impact on environment and reduced energy costs. By 2030, EVs are expected to globally share more than 80\% of vehicle market \cite{EV1}. However, consistent energy supply to increasing number of EVs is potentially a serious challenge for electricity grids \cite{EV2}. Recent research presents solutions to reduce the burden on main power grids by extensive exploitation of microgrids \cite{F1}-\cite{K1} but they require substantial setup and installation cost, and their supply is also limited \cite{micro1}. Optimizing charging schedules of EVs by determining peak demand hours \cite{Charging1}-\cite{Charging2} and adopting vehicle-to-grid (V2G) technology to store excess energy from EVs during off-peak hours have been widely suggested to ease off the electricity generation requirements \cite{opt1}. However, temporal prediction of demand from EVs is largely affected by outliers, such as an unusual incident and reduced travel during Covid-19 pandemic. Furthermore, battery storage systems are costly and experience degradation with time \cite{battery}. Apart from meeting rising energy demands, the grids must also adhere to environmental protection rules and restrictions on Carbondioxide (CO$_2$) emissions. Furthermore, achieving the net-zero goal in near future is another challenge.

A net-zero balance in a region means that it has at least as much energy available as it consumes with no or negligible CO$_2$ emissions \cite{NZ}. With rising number of EVs, it is extremely challenging to meet transport energy demands with renewable resources alone. One of the emerging solutions to meet high demands is prosumerism, where consumers offer their surplus energy in exchange of some incentive \cite{VNC1}. The energy trading systems involving EVs as sellers have been thoroughly analyzed using game theoretic approaches \cite{game1}. Optimized pricing strategies have also been proposed to support prosumerism \cite{pricing}. EVs can potentially play the role of prosumers if their batteries are sufficiently charged to sell a portion of their energy while also meeting their own demands. Therefore, the contribution of EVs in achieving net-zero is two-fold. They reduce emissions and can also act as energy suppliers to meet demands. However, one of the associated challenges is to find feasibility and the extent of practical applicability of prosumerism when EVs are traveling at high speeds on longer routes and their energy storage capabilities are limited. Secondly, distributed and scalable energy management systems are required with growing number of consumers and EVs \cite{Sp1} - \cite{Sp2}. Thirdly, EVs cannot maintain a net-zero balance alone if they are charged with energy generated from fossil fuels. Therefore, promoting renewable sources is equally important. The global policies to provide economic incentives for promoting clean energy is an additional benefit apart from the environmental friendly feature of renewable sources \cite{Zhang}.

Several renewable resources including wind, solar, geothermal, hydropower and bioenergy are exploited to provide clean energy with reduced emissions. Wind and solar energy are among the most abundant sources in the world which are not depleted by their usage. Rapid reduction in their installation cost has also been observed in the last decade \cite{Renew1}. Photovoltaic (PV) panels are used to convert solar energy into electricity. The output of a PV panel significantly depends upon solar irradiance at a particular time, which is largely affected by location, time and weather conditions \cite{solar1}. Wind energy also fluctuates due to variations in weather, wind speed and direction \cite{Wind1}. Therefore, forecasting energy output from PV and wind systems results with a certain degree of error. However, several machine learning (ML) models have been designed to achieve high accuracy and can attain a reasonable margin of uncertainty \cite{PV1} - \cite{Wind2}.

This paper proposes an energy management solution for main power grids involving EVs, wind and PV energy supporting the demand requirements of a transportation network. ML models are adopted to predict hourly demand and supply from EVs based on the routes and speeds, and supply from other distributed wind and PV energy sources. According to the expected demand, the grid controls the amount of supply from itself and EVs, and also pays incentive to EVs acting as prosumers. An optimization problem and a distributed incentive-based solution are formulated to reduce the cost paid by grid for achieving net-zero balance and reduce emissions occurred during electricity generation. In particular, the contributions of the paper are as follows
\begin{itemize}
	\item We investigate reliable ML models trained on publicly available datasets of wind and solar energy in the London to accurately predict demand and supply with least training times.
	\item We design a smart energy management solution to reduce the electricity generation burden of grid by utilizing supply offered by distributed renewable sources and EVs. An optimization approach to dynamically balance grid supply, cost and adequately meet energy demands in a massive transportation network consisting of various types of EVs is presented. The optimized solution also encourages reduction in emissions by introducing a penalty charge in the optimization solution of dynamically controlled grid supply and cost.
	\item We propose an incentive distribution mechanism to encourage EVs to become cooperative by offering their surplus energy and avoid selfish behavior. The incentive distribution game is analyzed using Prisoner's dilemma game.
	\item We theoretically derive the upper bounds of expected demand and surplus supply of EVs. Simulation results are evaluated considering a highway road with three types of EVs, i.e., cars, buses and lorries to evaluate the extent of reduction in grid load and cost. The optimum number of charging stations considering the maximum contribution from prosumers have also been discussed.
\end{itemize}

The rest of the paper is organized as follows. Section II discusses related works. The proposed solution of supply and demand prediction and optimization is described in Section III. Theoretical analysis and simulation results are presented in Section III and IV respectively, followed by conclusion in Section V.

\section{Related Works}
\begin{table*}[!t]
	\caption{Objectives of Energy Management and Optimization Strategies \label{table1}}
	\centering
	\begin{tabular}{|c|c|c|c|c|c|c|c|c|c|c|}
		\hline
		\textbf{Objective} & \textbf{\cite{Charging1}} & \textbf{\cite{Charging3}} & \textbf{\cite{Charging2}} & \textbf{\cite{opt1}} & \textbf{\cite{VNC1}} & \textbf{\cite{game1}} & \textbf{\cite{pricing}} & \textbf{\cite{pricing2}} & \textbf{\cite{MILP1}} & \textbf{Ours} \\
		\hline
		Reduce grid load  & \checkmark  &  \checkmark & \checkmark & \checkmark  &   &   & & \checkmark  & \checkmark  & \checkmark \\
		\hline
		Reduce grid cost &   &   &  &   &   &   &   &  & \checkmark  & \checkmark \\
		\hline
		Increase EV's utility  & \checkmark   & \checkmark  & \checkmark &   & \checkmark  & \checkmark  & \checkmark & \checkmark &   & \checkmark \\
		\hline
		Prosumerism &   &   &  & \checkmark  & \checkmark  & \checkmark  & \checkmark & \checkmark  & \checkmark  & \checkmark \\
		\hline
		Environmental Consideration &   &   &  &   & \checkmark  &   & &   & \checkmark  & \checkmark \\
		\hline
	\end{tabular}
\end{table*}
\begin{table*}
	\caption{Contributions of Related Works on Energy Optimization \label{tablecomp1}}
	\centering
	\begin{tabular}{|c|c|c|c|}
		\hline
		\textbf{Paper} & \textbf{Models for energy estimation}  & \textbf{Incentive Distribution Mechanism} & \textbf{Analysis} \\
		\hline
		\cite{Charging1} & Statistical & Knapsack problem based & Numerical (real data) \\
		\cite{Charging3} & None & None & Simulation \\
		\cite{Charging2} &  None & Differential game & Simulation \\
		\cite{opt1} &  None & None & Simulation \\
		\cite{VNC1} &  None & Stackelberg game & Simulation \\
		\cite{game1} &  None & Non-cooperative game & Simulation \\
		\cite{pricing} &  None & Bayesian game & Simulation \\
		\cite{pricing2} &  None & None & Simulation \\
		\cite{MILP1} &  Probabilistic & Penalty charge & Simulation \\
		{Ours} &  ML & Prisoner's dilemma game & Theoretical and simulation \\
		\hline
	 \end{tabular}
\end{table*}
This section firstly discusses related works of energy management and optimization strategies involving EVs, which is the main theme of this paper. Secondly, it reviews various energy estimation methods for wind, solar sources and EV demand, as they are essential for maintaining net-zero and optimize energy supply.
\subsection{Energy Management and Optimization Strategies}
Optimized strategies for effective energy management involving EVs have been proposed in literature with various objectives including demand and supply balance, maximizing grid profit, minimizing generation or charging price, and reducing number of charging times \cite{EV2}. The role of 5G base station (BS) for the cost-effective distribution of energy from grid is highlighted in \cite{5G}.

In \cite{Charging1} and \cite{Charging3}, charging schedule of EVs is optimized according to peak hours to minimize charging price for EVs. K-means clustering algorithm and Particle Swarm Optimization method is used in \cite{Charging1} and \cite{Charging3} respectively. Statistical prediction of energy is performed in \cite{Charging1} and Knapsack problem based ranking of EVs is proposed as an incentive. The effectiveness of the proposed solutions in \cite{Charging1} and \cite{Charging3} are evaluated through numerical data and examples. Charging times of EVs are also considered for pricing strategy by grid in \cite{Charging2}. Its objective is to reduce the grid load by utilization of differential game model for economic benefits. In \cite{pricing2}, grid load reduction is achieved by two approaches; bidirectional EV charging through V2G and time-based charging price management. Another approach to reduce grid load is battery storage systems which can store surplus energy from EVs through bidirectional V2G technology, which is presented in \cite{opt1}. It employs peak shaving strategy to reduce grid load.
 
The concept of environmental protection is introduced in \cite{MILP1} by imposing a penalty charge to users for per unit mass of CO$_2$ produced due to energy generation. A Mixed Integer Linear Programming (MILP) based optimization problem is solved to reduce energy generation costs and grid load through regulating charging schedule of devices in \cite{MILP1}. EVs are exploited as storage devices which can provide energy when needed. 

The concept of prosumers, i.e., EVs providing surplus energy, is proposed in some energy management solutions, where the main objective is usually maximizing the utility of EVs through game theoretic strategies \cite{VNC1} - \cite{pricing}. The theoretical probability of successful buyer and seller matching in energy trading is analyzed only in \cite{VNC1}. Blockchain is suggested to implement secure prosumerism and record energy transactions \cite{VNC1} - \cite{pricing}. 

Although prosumerism with the objective to reduce grid load is presented in literature but it lacks detailed analysis to evaluate its practical feasibility. This paper presents a consolidated solution comprising of incentives for EVs, penalty charges to reduce CO$_2$ emissions and theoretical analysis matched with numerical data to show the extent of demand covered by prosumerism. Contrary to the related solutions which have limited objectives, this paper covers multiple goals achieved through a single solution, as shown in Table~\ref{table1}. Also, Table~\ref{tablecomp1} shows the contributions presented in related works compared with this paper. 

\subsection{Energy Estimation and Prediction} 
\subsubsection{Wind and PV Energy} 
Forecasting of renewable energy output is essential to ensure a sufficient energy supply in a region. The varying weather conditions cause instability in the supply of wind and PV energies. The dataset used for predicting wind and PV energy includes spatial data, such as air temperature or wind speed in a region, sky images or temporal data such as amount of energy produced by wind or PV resource on a particular day and time of the year \cite{PVreview1}. Accurate energy estimation is challenging due to random nature of data. Most common estimation methods include statistical and ML models. Statistical models are usually derived from mathematical relationship between input and output data, which are often inaccurate due to non-linearity and complexity of the relationship \cite{PVreview2}. ML methods mostly include deep learning techniques such as Convolutional Neural Networks (CNN), Recurrent Neural Networks (RNN) or combination of both. Long Short Term Memory (LSTM) and gradient boosting are among the common RNN techniques to perform regression for predicting energy output \cite{solar1}, \cite{PVreview1}. Various datasets, ML models and accuracy metrics for predicting PV output are discussed in \cite{solar1} and a gradient boosting open-source framework XGBoost \cite{XGBoost} is resulted as one of the best energy estimation techniques. Also, a combination of CNN and LSTM model has been widely proposed as the efficient energy estimator for both wind and PV outputs \cite{windhybrid1} - \cite{PVhybrid2}. Due to high variations in datasets, techniques and models proposed in existing literature, the standard framework for energy estimation does not exist. Furthermore, current research focuses on the output accuracy of methods, whereas the training time of ML models and associated resource consumption in computing is not widely discussed. This paper compares the training time of various ML models for predicting wind and PV energy output.

\subsubsection{Electric Vehicles}
Energy demand prediction methods are considered as one of the source of motivation for road users to opt for EVs as they can easily plan their long journeys and stops according to the estimated energy consumption and demand on a route \cite{demand1} - \cite{demand2}. Furthermore, the demand prediction gives insight about the load forecast on a grid which can lead to effective energy management. Most of the energy demand prediction methods of EVs rely on ML methods \cite{demand1} - \cite{demand4}. In \cite{demand1} and \cite{demand3}, the history of EV charging at charging stations is used to predict demand and federated learning is employed to protect data privacy of each charging station. Time based data of energy demand at charging stations is also used for prediction in \cite{demand4} through a combination of LSTM and other regressive techniques. Forecasting demands of EVs on the basis of charging needs at charging stations results in a generic demand estimation at a particular location, irrespective of the individual EV's needs according to its specific planned route and consumption characteristics. In \cite{demand2}, transformer learning is used to determine the velocity of individual EVs according to their trajectory, route information and traffic flow characteristics. This method involves computation of acceleration or deceleration, and other EV specifications including battery capacity, mass and frontal area to mathematically calculate energy consumption on a route and predict demand. Similarly, a mathematical energy consumption algorithm based on EV specifications is also presented in \cite{demand5}, which can be used to predict demand according to a planned route.

\section{The Proposed Solution}
\begin{figure*}[!t]
	\centering
	\captionsetup{justification=centering}
	\includegraphics[scale=0.4]{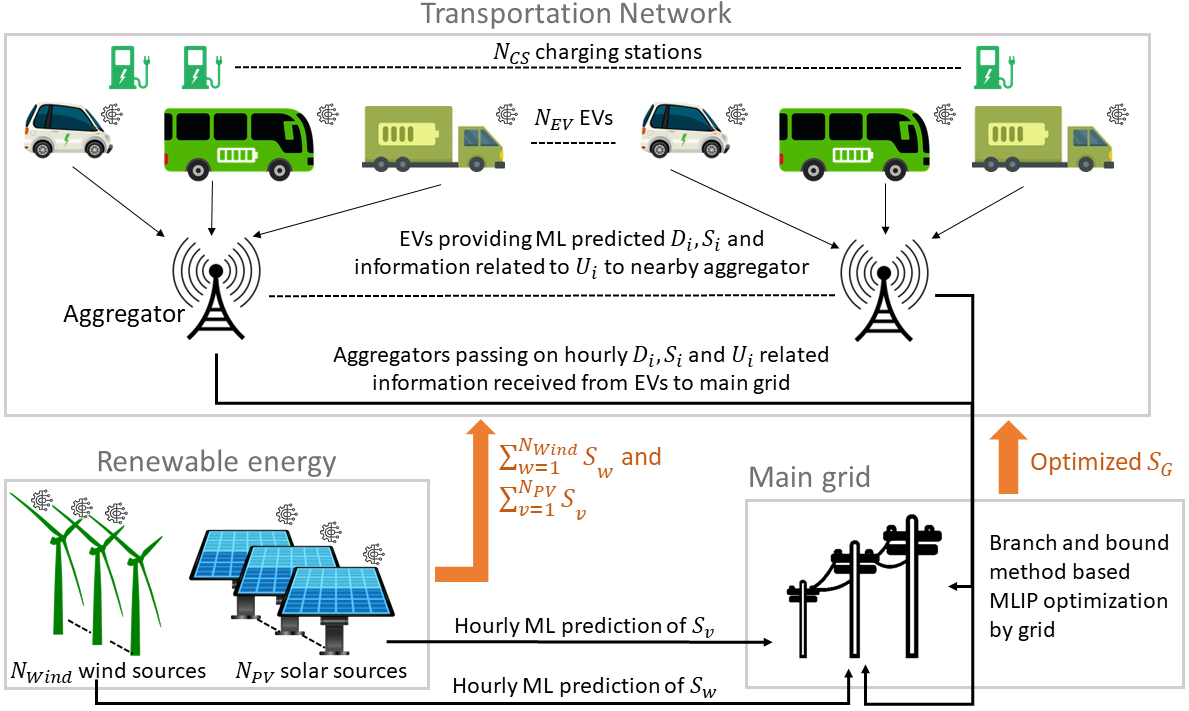}
	\caption{The proposed architecture of optimized energy management.}
	\label{fig_1}
\end{figure*}
\begin{table*}[!t]
	\caption{List of Key Notations \label{table2}}
	\centering
	\begin{tabular}{|c|c|c|c|}
		\hline
		\textbf{Notation} & \textbf{Definition} & \textbf{Notation} & \textbf{Definition}  \\
		\hline
		$N_{EV}$  & Number of EVs & $N_{CS}$  & Number of Charging Stations \\
		$N_{Wind}$ & Number of wind sources & $N_{PV}$  & Number of PV sources \\
		$S_i$ & Supply from EV $i$   &
		$D_i$ & Demand of EV $i$   \\
		$S_G$ & Supply from main grid &
		$S_w$, $S_v$ & Supply from wind, PV source \\
		$\hat{F}_i(k)$ & Traction force of EV $i$ at instant $k$ &
		$\hat{v_i}(k)$ & Velocity of EV $i$ at instant $k$ \\
		$\hat{a_i}(k)$ & Acceleration of EV $i$ at instant $k$ &
		$\hat{\tau}_i(k)$ & Torque of EV $i$ at instant $k$ \\
		$\hat{\omega}_i(k)$ & Speed of traction of EV $i$ at instant $k$ &
		$\eta_i(k)$ & Energy conversion efficiency of EV $i$ at instant $k$ \\
		$EP_i(k)$ & Electric power consumed by EV $i$ at instant $k$ &
		$l_{i,k}$ & Distance traveled by EV $i$ at instant $k$ \\
		$L_i$ & Spatial length of planned route of EV $i$ &
		$m_i$, $A_i$ & Mass, frontal area of EV $i$ \\
		$rw_i$ & Radius of wheels of EV $i$ &
		$gr_i$ & Effective gear ratio of EV $i$ \\
		$cr_i$, $cd_i$ & Rolling resistance, drag coefficient of EV $i$ &
		$MP_i$ & Motor power of EV $i$ \\
		$BC_i$ & Battery capacity of EV $i$ &
		$T_i$ & Total time of a route planned by EV $i$ \\
		$EC_i$ & Energy consumed by EV $i$ in time $T_i$ &
		$\theta$ & Road grade \\
		$\rho$ & Air density &
		$\alpha$ & Percentage of SOC which EV $i$ offers as $S_i$ \\
		$p_G$ & Price paid by grid to generate per unit $S_G$ &
		$SOC_{i,min}$, $SOC_{i,max}$ & Minimum and maximum SOC of EV $i$ \\
		$m_G$ & Mass of CO$_2$ produced during energy generation &
		$SOC_{i,c}$, $SOC_{i,r}$ & Current and remaining SOC of EV $i$ \\
		$PC$ & Penalty charge per unit CO$_2$ &
		$p_{EV}$ & Price paid by grid to EV for supplying $S_i$ \\
		$U_i$ & Utility of EV $i$ &
		$a$, $b$ & Energy cost factors \\
		$\beta$ & Incentive paid by regulation authorities to EVs &
		$\delta_i$ & Per unit energy consumption rate of EV $i$ \\
		$l_{S_i}$ & Distance traveled by EV $i$ to provide $S_i$ &
		$Loss$ & Transmission loss of supplying $S_G$ \\
		$Coop$ & Cooperative action &
		$Non-Coop$ & Non-cooperative action \\
		$N_{Coop}$ & Number of cooperative EVs &
		$N_{TH}$ & Threshold number of cooperative EVs \\
		$RC$ & Road Charge &
		$pr_x$ & Probability of occurrence of $x$ vehicle on road \\
		$\mu_{SOC}$ & Mean of $SOC_{i,r}$ & $\sigma^2_{SOC}$ & Variance of $SOC_{i,r}$\\ 	
		$L_R$ & Length of road & 	$\lambda$  & Arrival rate of EVs at charging station \\
		\hline
	\end{tabular}
\end{table*}
Fig.~\ref{fig_1} shows the proposed architecture of smart energy management solution for EVs. It is based upon a 5G-enabled Vehicle-to-everything (V2X) network where main grid, EVs, aggregators, wind and PV energy sources are able to communicate with each other. Table~\ref{table2} lists the notations used in this paper.
\subsection{System Architecture}
As shown in Fig.~\ref{fig_1}, the system architecture consists of sources providing supply to the transportation network including a main grid supplying $S_G$, and $N_{Wind}$ and $N_{PV}$ number of wind and PV energy sources each supplying an hourly energy amount of $S_w$ and $S_v$ respectively. The wind and PV sources predict their hourly $S_w$ and $S_v$ through ML models and share with main grid so it can adjust its $S_G$ accordingly. The transportation network includes 5G nodes serving as aggregators which are usually present in V2X networks to provide connectivity services to EVs, $N_{EV}$ number of EVs which can communicate with the aggregator. Each EV $i$ calculates its energy demand $D_i$ and the amount of surplus supply $S_i$ which it can offer for the next hour according to its ML based predicted velocity on a planned route and notifies the aggregator within its communication range. If an EV $i$ is offering $S_i$, it also shares parameters needed to compute its utility $U_i$, described later in this section. The main grid optimizes its $S_G$ to minimize its cost $C_G$ according to the accumulated demand and supply information received from the transportation network, and wind and PV sources. All EVs and their charging points are assumed to be incorporated with bidirectional chargers. The $S_i$ supplied by an EV $i$ can either be directly supplied to other EVs with energy demands or sent back to the grid via V2G technology through a bidirectional charger. There are $N_{CS}$ charging stations to facilitate energy trading among EVs and grid. 
\subsection{Energy Supply Prediction from Wind and PV Systems}
The proposed solution utilizes an open-source wind and PV energy output using meteorological satellite data and hourly simulation in the city of London \cite{ninja}. The data for predicting wind energy output consists of day, time, wind speed and wind direction at a particular location. The data for predicting PV energy output consists of day, time, air temperature, cloud opacity, dew point, precipitation, relative humidity, wind direction, wind speed, zenith and solar radiance. Four ML models including open-source CatBoost \cite{catboost}, XGBoost \cite{XGBoost}, Light Gradient Boosting Machine (LGBM) \cite{LGBM}, and a custom hybrid CNN-LSTM network are analyzed in terms of resulting Mean Absolute Error (MAE) and training time. Their comparative analysis is presented in Section IV.

\subsection{Energy Demand and Supply Prediction of EVs}
\begin{figure}[!t]
	\centering
	\captionsetup{justification=centering}
	\includegraphics[width=0.47\textwidth]{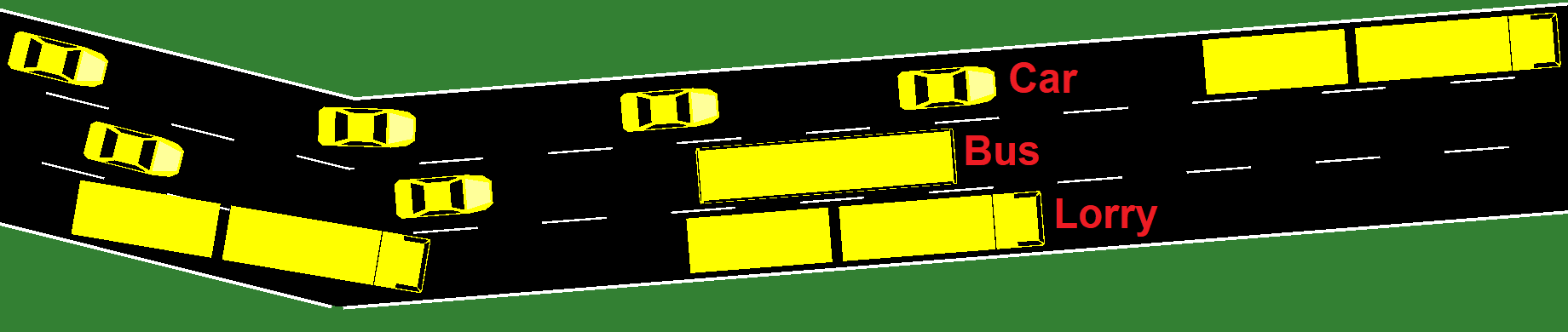}
	\caption{Traffic simulation in SUMO to collect data for energy demand and supply prediction of EVs.}
	\label{figSUMO}
\end{figure}
\begin{algorithm}[!t]
	\caption{$D_i$ and $S_i$ calculation by EV $i$.}\label{alg:alg1}
	\begin{algorithmic}[1]
		\Procedure{Determine $D_i$ and $S_i$} {}
		\State Get $SOC_{i,c}$.
		\State Estimate $SOC_{i,r}$ through (1) - (7).
		\State Set $\alpha$, $SOC_{i,min}$ and $SOC_{i,max}$.
		\If{$SOC_{i,r} \leq SOC_{i,min}$}
		\State $D_i = SOC_{i,max} - SOC_{i,c}$.
		\Else
		$\,D_i=0$.
		\EndIf
		\If{$SOC_{i,r} \geq SOC_{i,max}$}
		\State $S_i = \alpha \cdot SOC_{i,r}$.
		\Else
		$\,S_i=0$.
		\EndIf
		\State \textbf{return} $D_i$ and $S_i$.
		\EndProcedure
	\end{algorithmic}
	\label{alg1}
\end{algorithm}

The proposed solution of energy demand and supply prediction of EVs initially estimates the velocity changes of an EV $i$ on a planned route through an ML model. The dataset for velocity prediction of an EV $i$ includes its position, lane, angular direction of movement, maximum speed of road, and number of nearby vehicles. The proposed solution analyzes four ML models including CatBoost \cite{catboost}, XGBoost \cite{XGBoost}, LGBM \cite{LGBM}, and transformer learning \cite{demand2} to predict velocities of EVs with Root Mean Square Error (RMSE) as a performance metric. Then, a mathematical model to calculate energy consumption from velocity changes defined in \cite{demand2} is followed. 

For estimating energy consumption by an EV $i$, the total spatial length $L_i$ of its planned route is first divided into $K$ equal instances such that $L_i = \sum_{k=1}^{K} l_{i,k}$. After the velocity prediction at an instant $k$, i.e., $\hat{v}_i(k)$ by the ML model, its traction force $\hat{F}_i (k)$ is computed as
\begin{equation}
	\begin{aligned}
		\hat{F}_i(k) = &m_igsin\theta + m_igcos\theta \cdot cr_i \\ & +\bigg( \frac{\rho \cdot A_i \cdot cd_i \cdot \hat{v_i}(k)^2}{2} \bigg) + m_i \hat{a_i}(k),
	\end{aligned}
\end{equation}
where $g$ is the gravitational acceleration, $\theta$ is the road grade, $\rho$ is the air density, and $m_i$, $A_i$, $cr_i$, $cd_i$ and $\hat{a_i}(k)$ is the mass, frontal area, rolling resistance coefficient, aerodynamic drag coefficient and acceleration of EV $i$ at instant $k$ respectively. $\hat{a_i}(k)$ is defined as
\begin{equation}
	\hat{a}_i(k) = \frac{\hat{v}_i(k)^2 - \hat{v}_i(k-1)^2}{2l_{i,k}}.
\end{equation}
The torque $\hat{\tau}_i(k)$ and speed of traction $\hat{\omega}_i(k)$ are defined in \cite{force} as
\begin{equation}
	\hat{\tau}_i(k) = \frac{\hat{F}_i(k) rw_i}{gr_i}, \, \, \,	\hat{\omega}_i(k) = \frac{\hat{v}_i(k) gr_i}{rw_i},
\end{equation}
where $rw_i$ and $gr_i$ is the radius of wheel and effective gear ratio of EV $i$ respectively. The energy conversion efficiency from electrical to mechanical energy at instant $k$, $\eta_i(k)$, is
\begin{equation}
	\eta_i(k) = \frac{\hat{\tau}_i(k) \hat{\omega}_i(k)}{MP_i}, 
\end{equation}
where $MP_i$ is the motor power of EV $i$. The electric power consumed at instant $k$ is
\begin{equation}
	EP_i(k) = \frac{\hat{F}_i(k) \hat{v}_i(k)}{\eta_i(k)}.
\end{equation}
The total energy consumed on a planned route is then given as
\begin{equation}
	EC_i = \sum_{k=1}^{K} EP_i(k) \cdot T_i,
\end{equation}
where $T_i$ is the estimated total time to travel on the planned route of EV $i$. The remaining State of Charge (SOC) after its journey is
\begin{equation}
	SOC_{i,r} = SOC_{i,c} - EC_i,
\end{equation}
where $SOC_{i,c}$ is its current SOC. An EV $i$ calculates its demand $D_i$ and available $S_i$ as
\begin{equation}
	D_i  = \begin{cases}
		SOC_{i,max} - SOC_{i,c}, & SOC_{i,r} \leq SOC_{i,min}, \\
		0,  & \text{Otherwise,} \\
	\end{cases}
\end{equation}
and
\begin{equation}
	S_i  = \begin{cases}
		\alpha \cdot SOC_{i,r} , & SOC_{i,r} \geq SOC_{i,max}, \\
		0,  & \text{Otherwise,} \\
	\end{cases}
\end{equation}
where $ SOC_{i,min}$ and $ SOC_{i,max}$ are the minimum and maximum SOC limits respectively which can be set according to the battery capacity $BC_i$ of each EV $i$ and $\alpha$ denotes the percentage of its $SOC_{i,r}$ which it offers as $S_i$. Algorithm~\ref{alg1} defines whether an EV $i$ needs $D_i$ or can offer $S_i$ during its planned route.

\subsection{Optimization Problem}
\begin{figure}[!t]
	\centering
	\captionsetup{justification=centering}
	\includegraphics[width=0.47\textwidth]{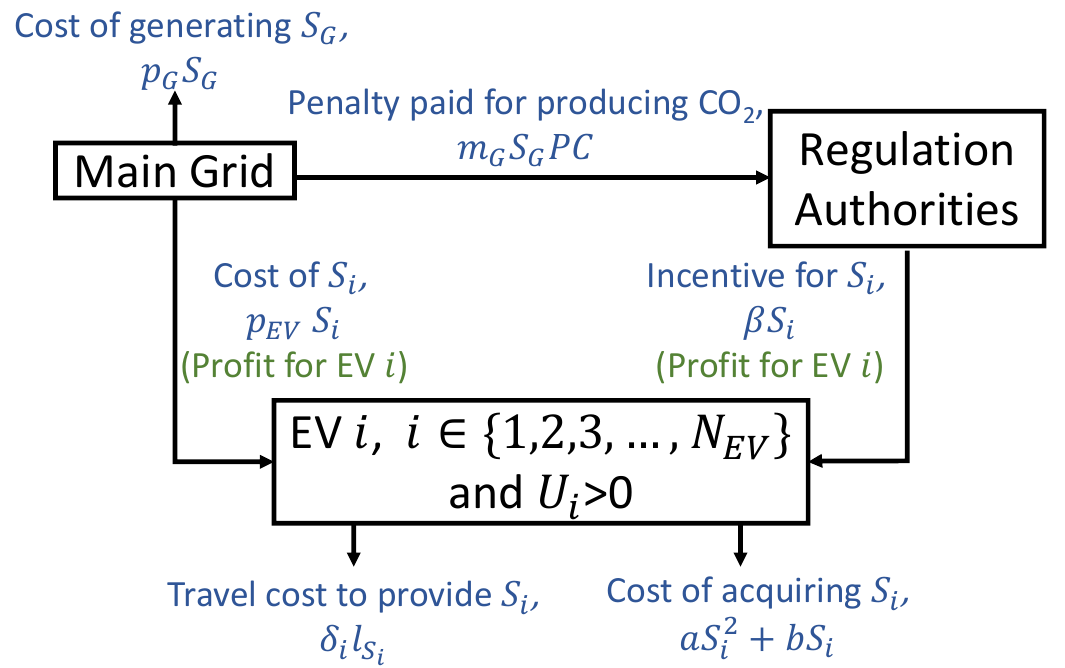}
	\caption{Costs and profits of main grid and EVs.}
	\label{figcost}
\end{figure}
\begin{algorithm}[!t]
	\caption{Cost Minimization by Main Grid.}\label{alg:alg2}
	\begin{algorithmic}[1]
		\Procedure{MILP} {}
		\State Estimate $\overline{S_G}$ and $p_G$.
		\State Set $p_{EV}$.
		\State Get $\beta$ from regulation authorities.
		\State Get $\sum_{w=1}^{N_{Wind}} S_w$ and $\sum_{v=1}^{N_{PV}} S_v$ from wind and PV sources respectively.
		\State Get $S_i$, $D_i$, $a$, $b$, $\delta_i$, $l_{S_i}$ of EV $i$  $\forall \, i \, \epsilon \, \{1, 2, 3,....,N_{EV} \}$ from aggregator.
		\While {$C_G \neq min(C_G)$}
		\State Optimize $S_G$ and $S_i$ through branch and bound algorithm.
		\State Ensure C1, C2 and C3 are true.
		\EndWhile
		\State \textbf{return} Optimized values $S_G$ and $S_i$.
		\EndProcedure
	\end{algorithmic}
	\label{alg2}
\end{algorithm}
The optimization problem aims to minimize the cost of electricity generation by main grid. The incurring grid costs for supplying $S_G$, paying for $S_i$, and cost and profit of EV $i$ are shown in Fig.~\ref{figcost}. The total cost $C_G$ for a grid to meet demands of a transportation network is
\begin{equation}
	C_G = p_G S_G + m_G S_G PC + p_{EV} \sum_{i=1}^{N_{EV}} S_i,
	\label{eqcost}
\end{equation}
where $S_G$ is the energy supplied by grid, $p_G$ is the cost of generating $S_G$, $m_G$ is the per unit mass of CO$_2$ released to produce $S_G$, $PC$ is the penalty charge paid by grid to regulation authorities for producing per unit CO$_2$ and $p_{EV}$ is the price paid by grid to each EV $i$ for selling $S_i$. The utility of EV $i$ for selling $S_i$ is
\begin{equation}
	U_i = (p_{EV} + \beta )S_i  - \delta_i l_{S_i} - aS_i^2 - bS_i ,
\end{equation}
where $\beta$ is the incentive paid by the regulation authorities to EVs for acting as prosumers. It can be offered as a benefit, for example, discount in annual tax or paid parking in an area. $l_{S_i}$ is the distance that EV $i$ has to travel to provide $S_i$, $\delta_i$ is the per unit energy consumption rate, $aS_i^2 + bS_i$ represents the cost of acquiring $S_i$, where $a>0$ and $b>0$ are the energy cost factors \cite{VNC1}. The cost function is assumed to be quadratic corresponding to linear decreasing marginal benefit \cite{Q1} - \cite{Q2}. An EV $i$ can choose $l_{S_i}$ itself by selecting one of the charging stations available within the vicinity of its route. However the nearest charging station results in least travel cost.

To maintain demand-supply balance of a transportation network for net-zero, it is must that
\begin{equation}
S_G + \sum_{w=1}^{N_{Wind}} S_w + \sum_{v=1}^{N_{PV}} S_v + \sum_{i=1}^{N_{EV}} S_i - \sum_{i=1}^{N_{EV}}D_i - Loss \geq 0 ,
\label{netzero}
\end{equation}
where $Loss$ is the transmission loss occurred while providing $S_G$. After receiving demand and supply information from a transportation network, the main grid solves the following optimization problem.

\textbf{Problem 1: }
\begin{mini*}|s|
	{S_G, S_i}{ C_G \, \forall \, i \, \epsilon \, \{1, 2, 3,....,N_{EV} \},}
	{}{}
	\addConstraint{\text{C1: }S_G  \leq \overline{S_G} }
	\addConstraint{\text{C2: }U_i > 0\, \text{if} \, S_i>0 \, \, \forall \, i \, \epsilon \, \{1, 2, 3,....,N_{EV} \} }
	\addConstraint{\text{C3: }(\ref{netzero}) .}{}
\end{mini*}
where constraint C1 represents that the grid is able to supply only energy up to a certain upper limit, i.e., $\overline{S_G}$, C2 represents that each EV $i$ sells $S_i$ to gain positive $U_i$ only and C3 represents net-zero balance between total demands and supplies. Algorithm~\ref{alg2} defines the solution carried out by main grid to solve Problem 1 by utilizing branch and bound method based on MILP framework. The motivation to exploit branch and bound method is its suitability with the combinatory nature of Problem 1, where there is a finite set of available $S_i$. The branch and bound method creates a search space tree with each node containing combination of information and attempts to find an optimal solution. This method offers a lower time complexity than other MILP solutions \cite{bb}.
\section{Theoretical Analysis}

\begin{table}
	\caption{Payoff matrix of EV $i$ \label{tablemat}}
	\centering
	\begin{tabular}{|c|c|c|}
		\hline
		{Action}	& 	$Coop$ & $Non-Coop$ \\
		\hline
		{Less than $N_{TH}$ choose $Coop$} & $U_i$ & $-RC$ \\
		\hline 
		{$N_{TH}$ or more choose $Coop$} & ${-RC}/{N_{Coop}}$ & ${-RC}/{N_{Coop}}$ \\
		\hline
	\end{tabular}
\end{table}
\subsection{Game Theoretic Analysis of Payoff for EV}
We analyze the strategy of an EV $i$ to offer $S_i$ in exchange of $U_i$ through Prisoner's dilemma game. In a game theoretic context, the payoff of an EV player $i$ depends upon its own action as well as the action of other EV players.
\paragraph{Actions} In this game, every EV player $i$ has two possible actions. It can either be cooperative ($Coop$) by offering $S_i$ or non-cooperative ($Non-Coop$) if it does not have surplus $S_i$ or choose to be selfish.
\paragraph{Payoff} The payoff matrix in Table~\ref{tablemat} shows the utilities of EV players according to their actions, where $RC$ represents the road charge that every player has to pay while passing through the road. To encourage prosumerism, the road charge reduces to $\frac{RC}{N_{Coop}}$ if $N_{Coop} \geq N_{TH}$, where $N_{Coop}$ denotes number of EVs playing cooperatively and $N_{TH}$ is an arbitrary threshold number of players. The road charge payoff reduces for both cooperative and non-cooperative players when $N_{Coop} \geq N_{TH}$ to ensure fairness among all EVs because a non-cooperative player is not necessarily selfish and may not have enough $S_i$ available to offer.

We prove the following propositions to analyze the motivating capability of the game for the player $i$ to take cooperative action.

\textbf{Proposition 1:} Playing cooperative ($Coop$) is the best response action of a player $i$ if $U_i \geq 0$ and $RC \geq 0$.
\\ \textit{Proof:} If $U_i \geq 0$ then $U_i > - \frac {RC}{N_{Coop}} > -RC$. A player $i$ will always get a payoff greater than or equal to $- \frac {RC}{N_{Coop}}$ if it chooses $Coop$. On the other hand, it will always get a payoff less than or equal to $- \frac {RC}{N_{Coop}}$ if it chooses $Non-Coop$. Therefore, $Coop$ is the best response action of player $i$ irrespective of the actions of other players when $U_i \geq 0$ and $RC \geq 0$. Hence, a player $i$ will be motivated to offer $S_i$ whenever available. \qed

\textbf{Proposition 2:} If a player $i$ has $S_i$ to offer, it will not conspire to act selfishly by choosing $Non-Coop$, even if it maliciously finds out that $N_{TH}$ or more choose $Coop$, only when $U_i \geq RC - \frac{RC}{N_{Coop}}$.
\\ \textit{Proof:} Let $q$ be the probability that less than $N_{TH}$ players choose $Coop$, irrespective of the knowledge of player $i$. Let $q_c$ be the probability that player $i$ conspires to find out that another player will choose $Coop$. Then the probability that player $i$ knows that $N_{TH}$ players will choose $Coop$ is $q^{N_{TH}}$. The expected payoff sum $E(Payoff)$ is
\begin{equation}
	\begin{aligned}
		& E (Payoff)  =  q'^{N_{TH}} \Big( - \frac {RC}{N_{Coop}} \Big) \\ & + (1 - q'^{N_{TH}}) \Bigg( q (U_i - RC) + (1-q) \Big (- \frac{RC}{N_{Coop}} \Big ) \Bigg), 
	\end{aligned}
\end{equation}
which reduces to
\begin{equation}
	\begin{aligned}
		E	(Payoff) & =  q \Big (U_i - RC + \frac{RC}{N_{Coop}} \Big) - \frac{RC}{N_C} \\ & - qq'^{N_{TH}} \Big(U_i - RC + \frac{RC}{N_{Coop}} \Big).
	\end{aligned}
\end{equation}
We want $E(Payoff) \leq E(Payoff')$ to prevent selfishness, where $Payoff'$ is the payoff of a player taking action without any conspiracy or knowledge of other players' actions, i.e.,
\begin{equation}
	E	(Payoff') =  q(U_i - RC) + (1-q) \bigg ( - \frac{RC}{N_{Coop}} \bigg ) . 
\end{equation}
If $E	(Payoff) \leq E	(Payoff')$,
\begin{equation}
	- qq'^N_{TH} \bigg (U_i - RC + \frac{RC}{N_{Coop}} \bigg) \leq 0,
\end{equation}
or
\begin{equation}
	U_i - RC + \frac{RC}{N_{Coop}} \geq 0,
\end{equation}
i.e., 
\begin{equation}
	U_i \geq RC - \frac{RC}{N_{Coop}} .
	\label{eqU}
\end{equation}
\qed
\subsection{Theoretical Bounds of Supply and Demand}
Let $SOC_{i,r}$ follow lognormal distribution with mean $\mu_{SOC}$ and variance $\sigma_{SOC}^2$ \cite{SOC}. The probability that  $SOC_{i,r} \geq SOC_{i,max}$ is
\begin{equation}
	\begin{aligned}
	& Pr(SOC_{i,r} \geq SOC_{i,max}) \\ & = 1 - \frac{1}{2} \Bigg(1+erf \Big( \frac {log(SOC_{i,max})-\mu_{SOC}} {\sqrt{2 \sigma_{SOC} }} \Big) \Bigg),
	\end{aligned}
\end{equation}
and the probability that  $SOC_{i,r} \leq SOC_{i,min}$ is
\begin{equation}
		\begin{aligned}
	& Pr(SOC_{i,r} \leq SOC_{i,min}) \\ & =  \frac{1}{2} \Bigg(1+erf \Big( \frac {log(SOC_{i,min})-\mu_{SOC}} {\sqrt{2 \sigma_{SOC} }} \Big) \Bigg).
	\end{aligned}
\end{equation}
Let $pr_x$ denote the probability of occurring car, bus and lorry respectively on road at a certain instant, where $x$ represents car, bus or lorry. The upper bounds of $S_i$ and $D_i$ are
\begin{equation}
	S^{UB} (x) = N_{EV} pr_x Pr(SOC_{i,r} \geq SOC_{i,max}) \cdot  \alpha SOC_{i,r} ,
\end{equation}

\begin{equation}
	\begin{aligned}
	D^{UB} (x)  & =  N_{EV} pr_x Pr(SOC_{i,r} \geq SOC_{i,max}) \\& \cdot (SOC_{i,max} - SOC_{i,c}) .
\end{aligned}
\end{equation}
\subsection{Expected Number of Prosumers per Charging Station}
Assuming that $l_{S_i}$ is the average distance of an EV $i$ to its nearest charging station and the number of charging stations are uniformly distributed on a road of length $L_R$. Then $N_{CS} = \lfloor \frac{L_R}{l_{S_i}} \rfloor$ is the total number of charging stations. The expected number of EVs supplying energy per charging station can be modeled using Poisson distribution \cite{FL1}
\begin{equation}
	E(N_{EV}^{Supp}) = \sum_{k=0}^{N_{CS}} k e^{- \lambda^{Supp} / N_{CS} } \frac{ (\lambda^{Supp}/{N_{CS}})^k}{k!},
\end{equation}
where $\lambda^{Supp}$ is the arrival rate of seller EVs. For buyer EVs, $\lambda^{Dem}$ is large and a normal approximation to Poisson distribution can be applied with mean $\lambda^{Dem}$ \cite{NP}. Therefore the expected number of demanding EVs per charging station is $E(N_{EV}^{Dem}) = \lambda^{Dem} / N_{CS}$.

\begin{figure*}%
	\centering
	\captionsetup{justification=centering}
	\subfloat[\fontfamily{ptm} \selectfont \small Wind energy prediction]{{\includegraphics[width=5.66cm]{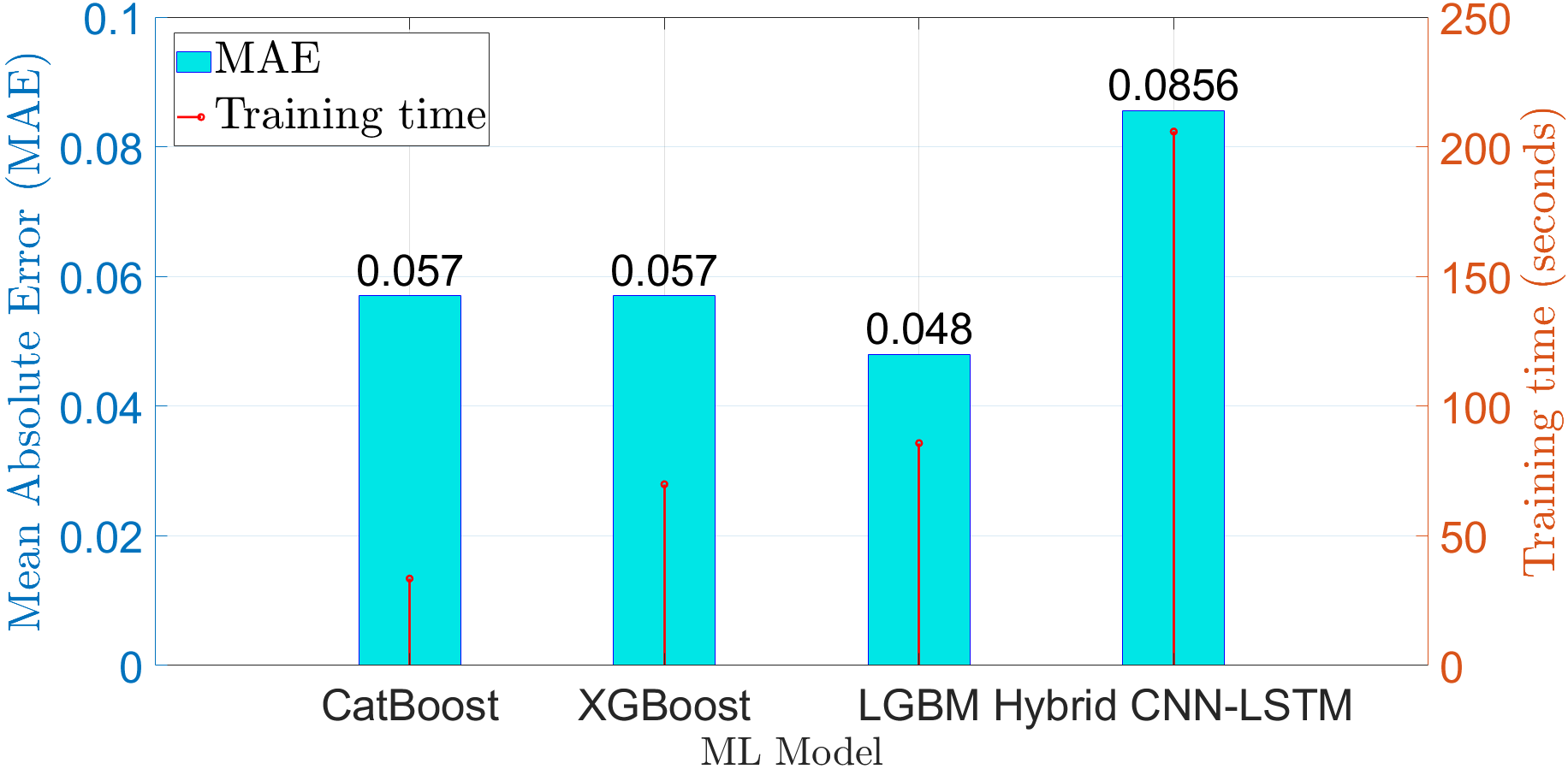} }}%
	\quad
	\subfloat[\fontfamily{ptm} \selectfont \small PV energy prediction]{{\includegraphics[width=5.66cm]{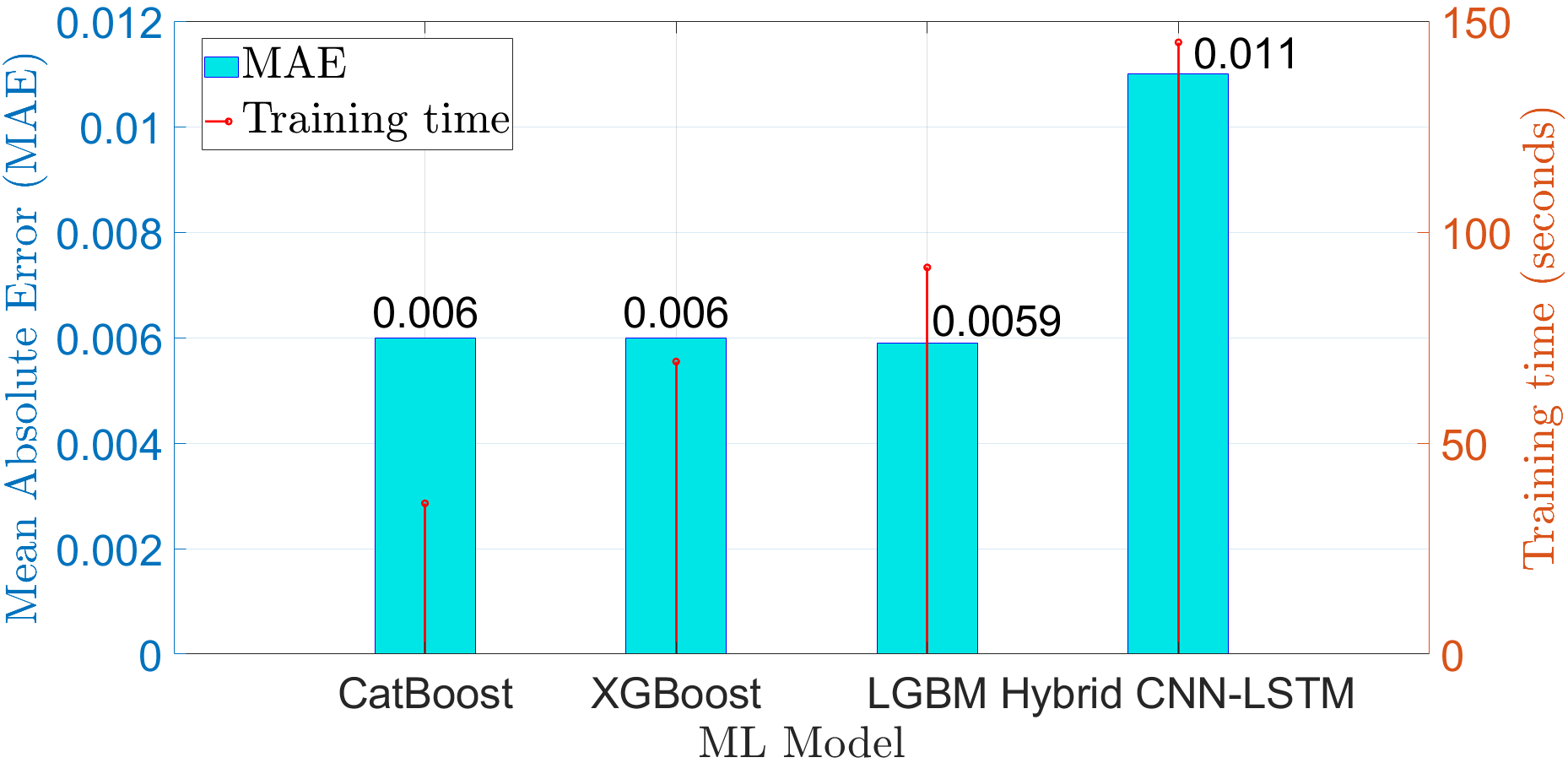} }}%
	\quad
	\subfloat[\fontfamily{ptm} \selectfont \small EV's velocity prediction]{{\includegraphics[width=5.66cm]{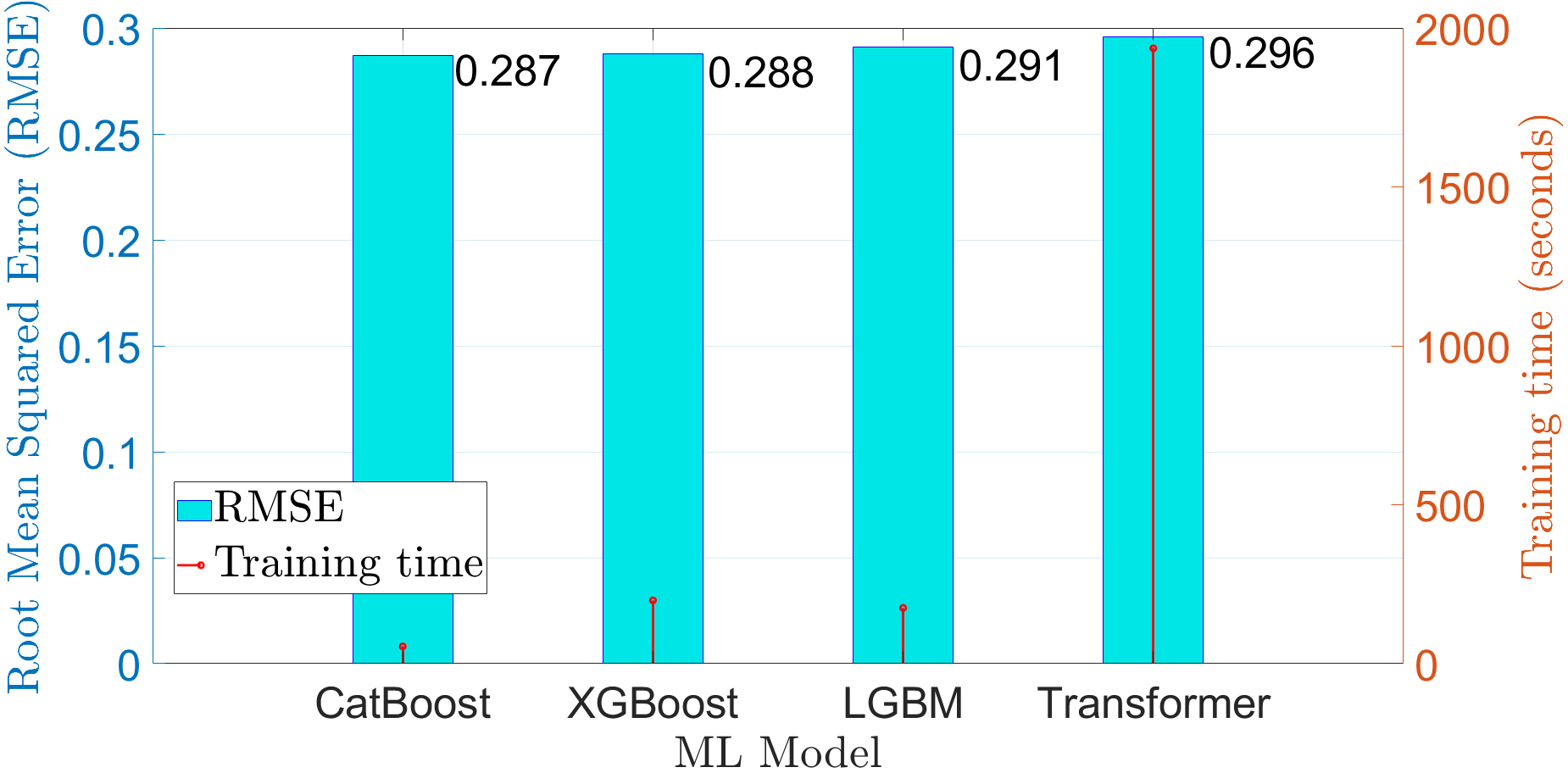} }}%
	\caption{Performance comparison of ML models.}
	\label{figMLa}
\end{figure*}
\begin{figure}%
	\centering
	\captionsetup{justification=centering}
	\includegraphics[width=0.47\textwidth]{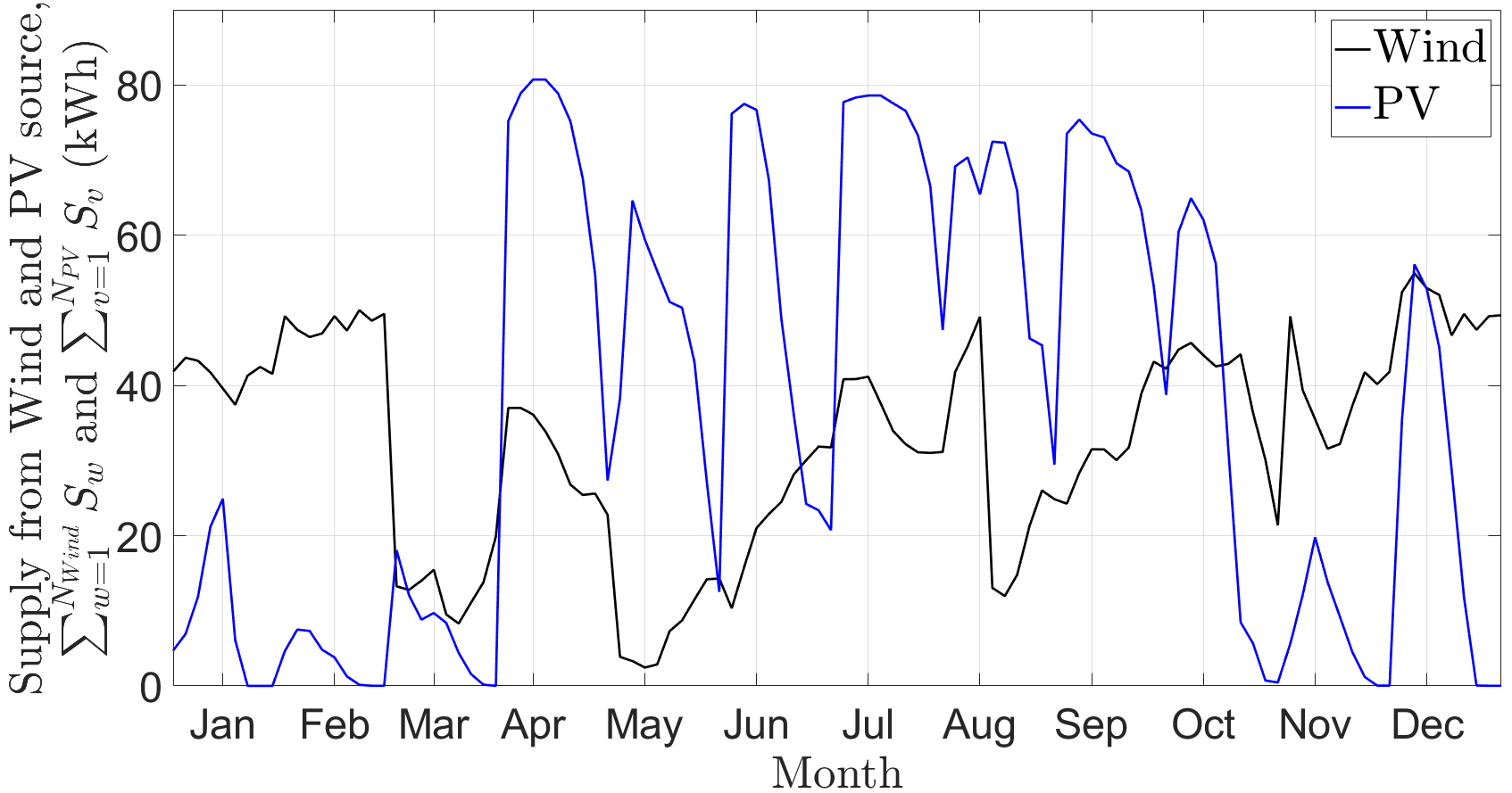}
	\caption{Wind and PV energy output, $N_{wind} = N_{PV} = 100$.}
	\label{figrenewCont}
\end{figure}
\section{Results and Discussion}
\begin{figure*}%
	\centering
	\captionsetup{justification=centering}
	\subfloat[\fontfamily{ptm} \selectfont \small Average Utility of EV $i$]{{\includegraphics[width=8.5cm]{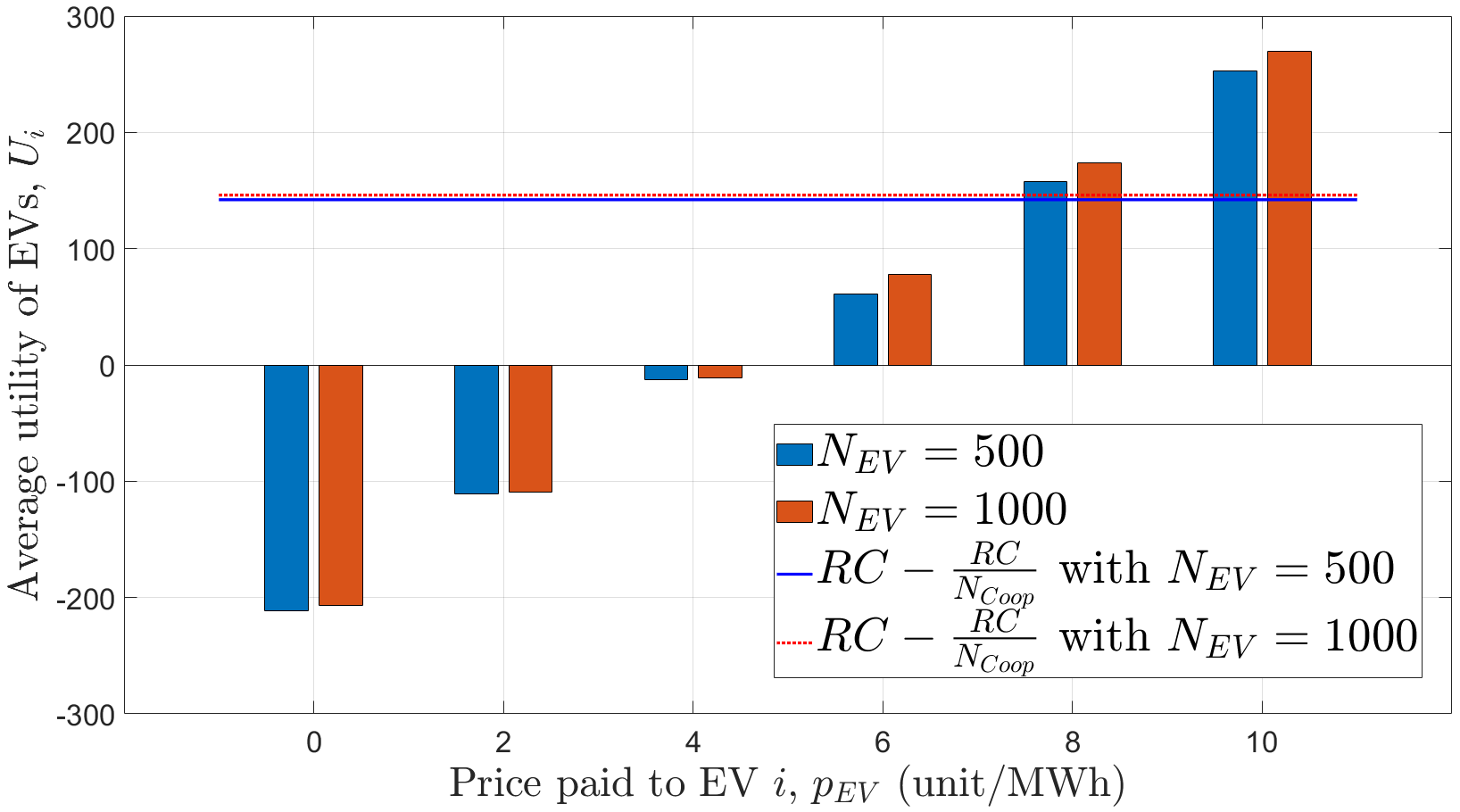} }}%
	\quad
	\subfloat[\fontfamily{ptm} \selectfont \small Comparison of offered and optimized $S_i$, $m_G$ = 0.05]{{\includegraphics[width=8.5cm]{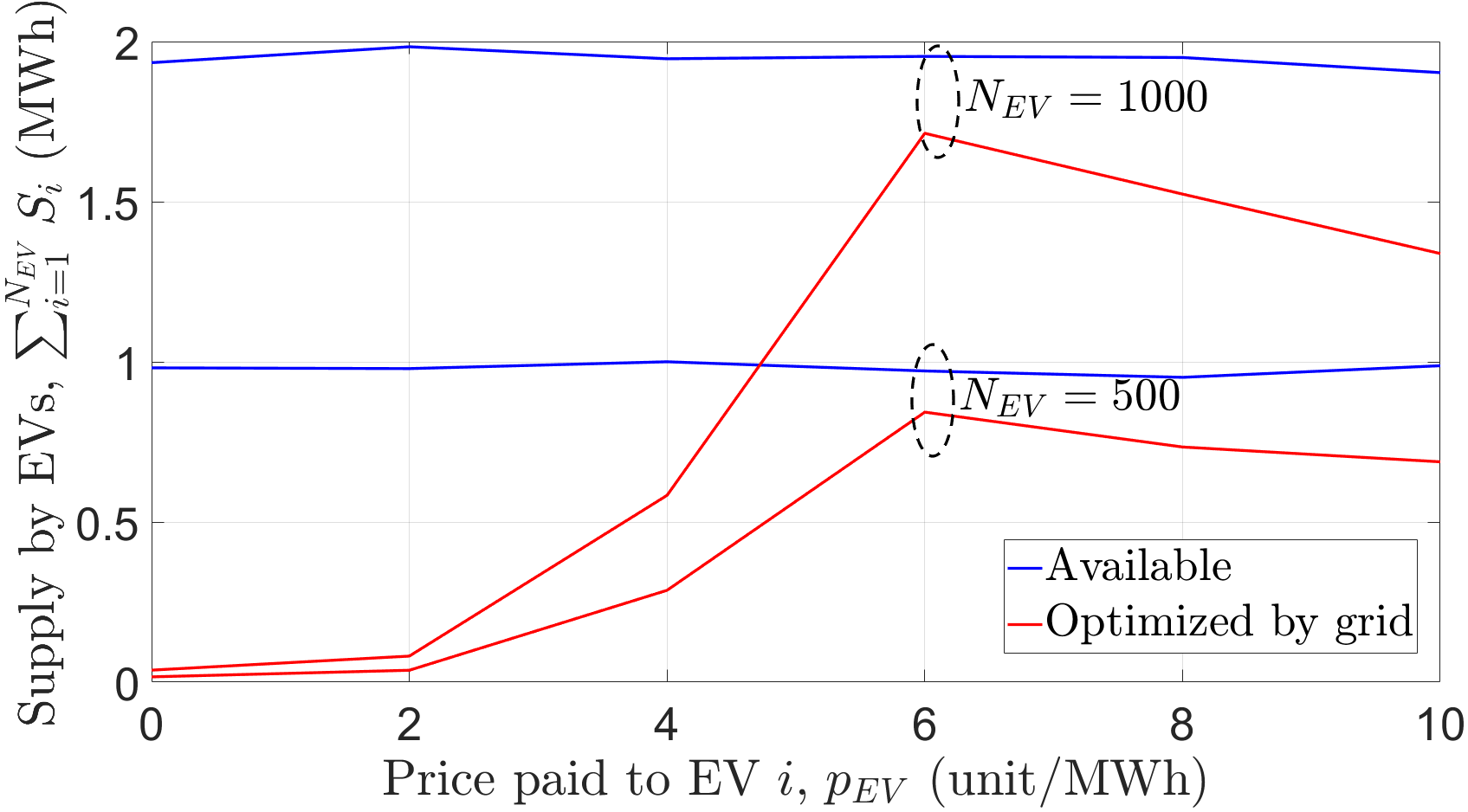} }}%
	\caption{Effect of $p_{EV}$ upon supplying $S_i$ by EV $i$.}
	\label{figpEV}
\end{figure*}
\begin{figure}%
	\centering
	\captionsetup{justification=centering}
	\includegraphics[width=0.47\textwidth]{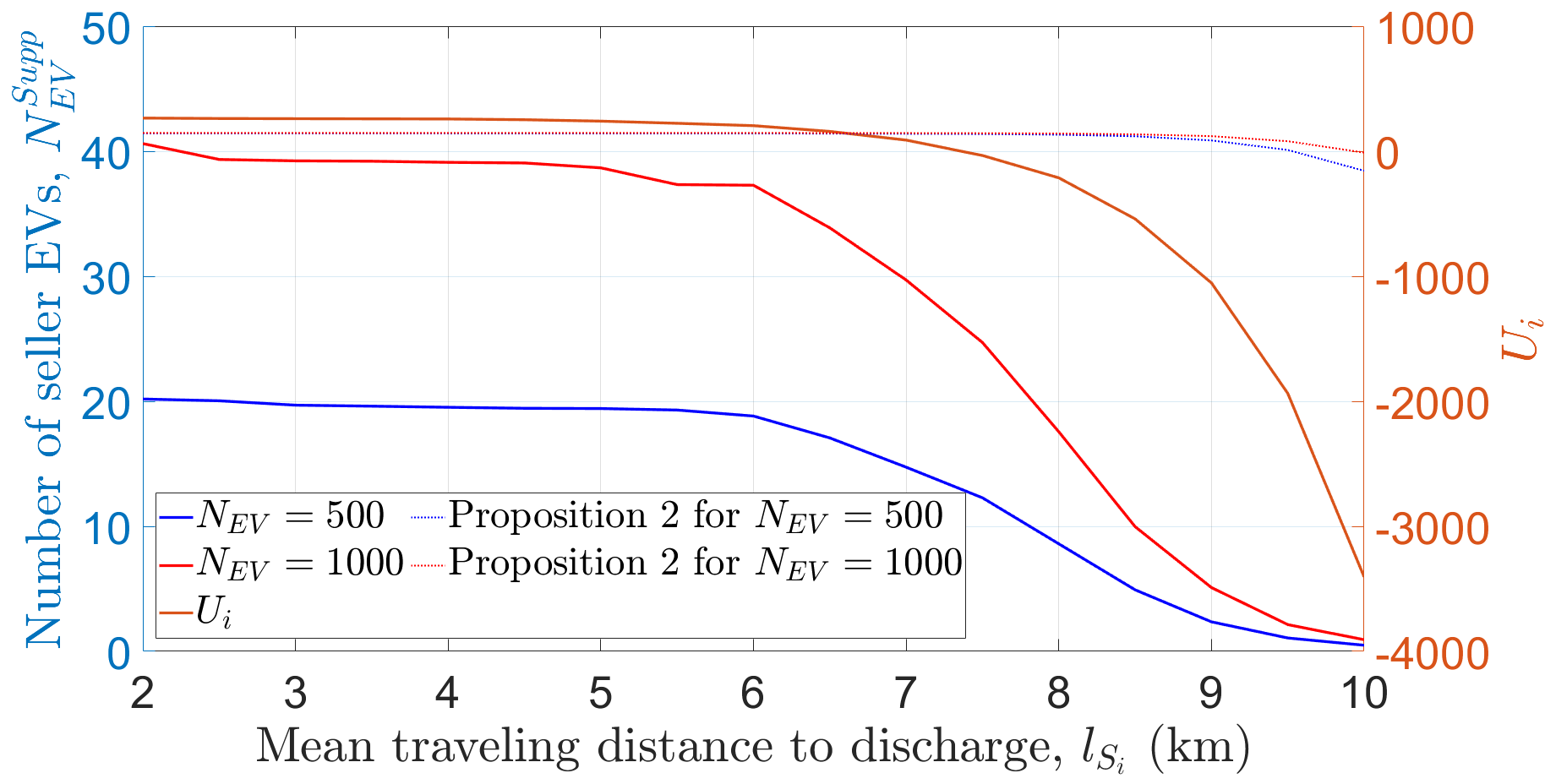}
	\caption{$U_i$ and $N_{EV}^{Supp}$ with respect to $l_{S_i}$.}
	\label{figchargeUi}
\end{figure}
\begin{figure}%
	\centering
	\captionsetup{justification=centering}
	\includegraphics[width=0.47\textwidth]{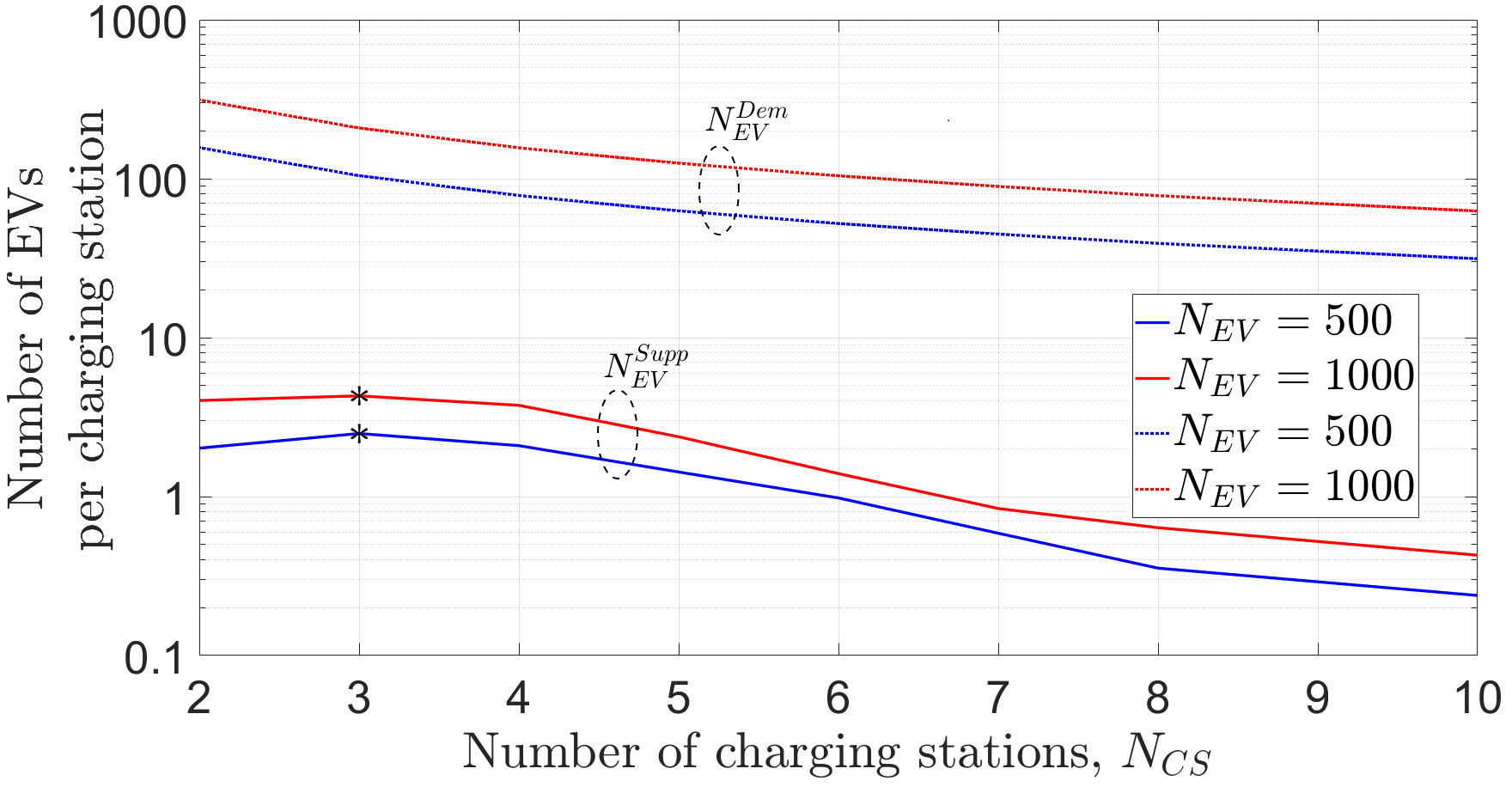}
	\caption{$E(N_{EV}^{Supp})$ and $E(N_{EV}^{Dem})$ with respect to $N_{CS}$.}
	\label{figNcharge}
\end{figure}
\begin{figure}%
	\centering
	\captionsetup{justification=centering}
	\includegraphics[width=0.47\textwidth]{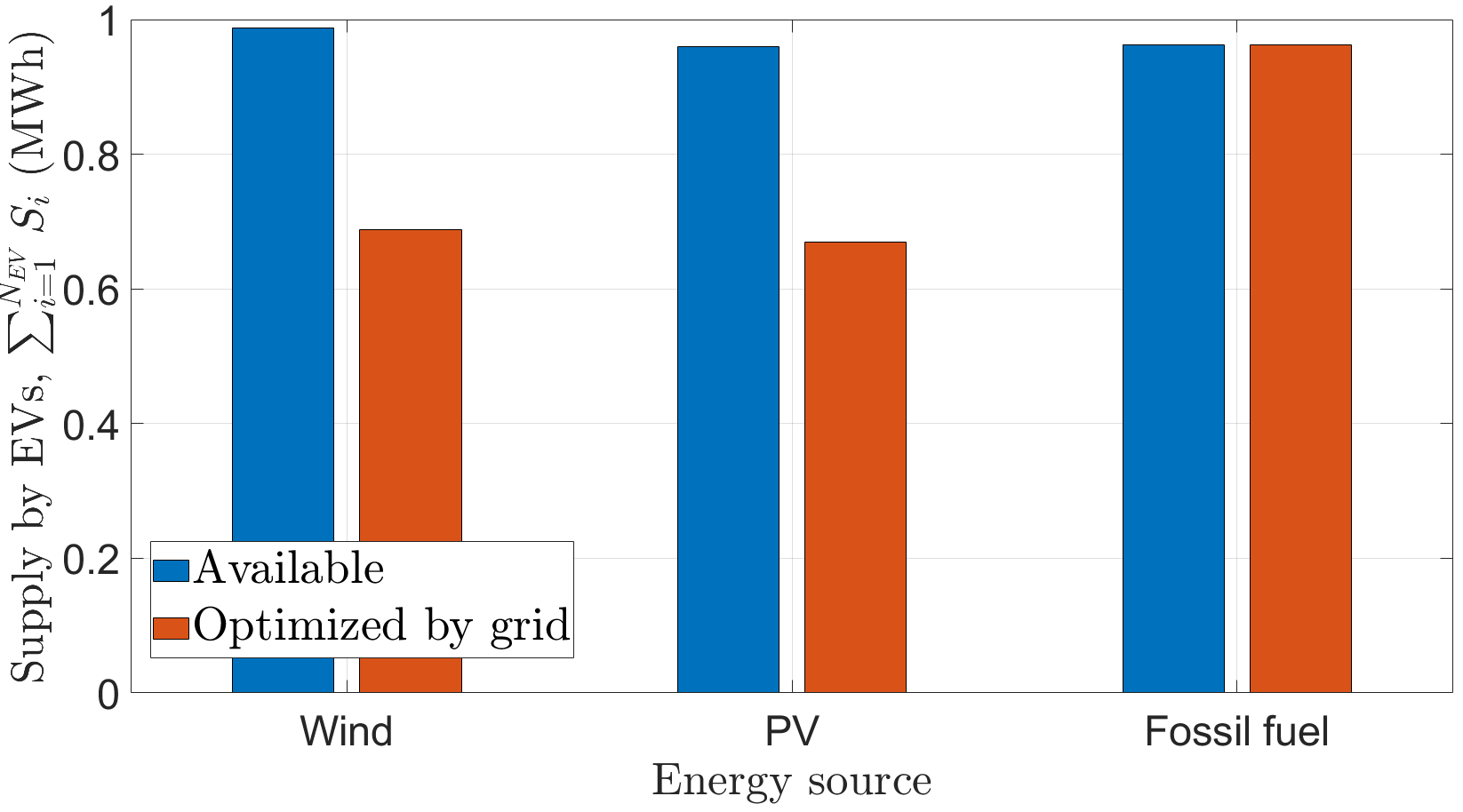}
	\caption{Effect of grid energy source on prosumerism, $N_{EV}=500$ and $p_{EV}=10$.}
	\label{figenergysource}
\end{figure}
\begin{table}
	\caption{Simulation Parameters \label{tablesim1}}
	\centering
	\begin{tabular}{|c|c|c|c|}
		\hline
		\textbf{Parameter} & \textbf{Value} & \textbf{Parameter} & \textbf{Value} \\
		\hline
		$N_{EV}$ & [500, 2000] & $N_{Wind}$, $N_{PV}$  & [50, 100]  \\ \hline
		$\theta$ & 6\% & $\rho$ & 1.28 kg/m$^3$ \\ \hline
		$SOC_{i,min}$ & 20\% of $BC_i$  & $SOC_{i,max}$ & 80\% of $BC_i$ \\ \hline
		$\alpha$ & 20\% of $SOC{i,c}$ & $p_G$ & 12 \\ \hline
		$PC$ & 10 & $Loss$ & 2\% of $S_G$ \\ \hline
		$p_{EV}$ & [0, 10] & $\beta$ & 10 \\ \hline
		$\delta_i$ & 0.1 & $RC$ & 150 \\ \hline
		$a, b$ & 0.01, 0.1 & $L_R$ & 20\,km \\ \hline
		$m_G$ & \multicolumn{3}{c|}{Wind: 0.05, PV: 0.088, Fossil fuel: 0.786 kg} \\
		\hline
	\end{tabular}
\end{table}
\begin{table}
	\caption{EV Specifications \label{tableEV}}
	\centering
	\begin{tabular}{|c|c|c|c|}
		\hline
		\multirow{2}{*}{\textbf{Parameter}} & \multicolumn{3}{c|} {\textbf{Value}} \\
		\cline{2-4}
		& \textbf{Car} & \textbf{Bus} & \textbf{Lorry} \\ \hline
		$cr_i$ & 0.01  & 0.08  & 0.011  \\ 
		$cd_i$ & 0.28 & 0.6 & 0.8 \\ 
		$m_i$ (kg) & 1619  & 2375 & 3556 \\ 
		$A_i$ (m$^2$) & 2.56 & 2.56 & 5.98 \\ 
		$rw_i$ (m) & 0.2 & 0.28 & 0.28 \\
		$gr_i$ & 7.94 & 3.98 & 3.73 \\ 
		$MP_i$ (kW) & 110 & 200 & 220 \\ 
		$BC_i$ (kWh) & 40 & 320 & 112 \\
		$p_x$ & 60\% & 40\% & 40\% \\
		$\mu_{SOC}$ (kWh) & 5 & 50 & 10 \\ 
		$\sigma_{SOC}$ & \multicolumn{3}{c|}{0.1} \\
		\hline
	\end{tabular}
\end{table}
\subsection{Simulation Setup}
We analyze the performance of the proposed solution using Python and relevant open-source libraries including sci-kit, PuLP, catboost, xgboost, lightgbm and tensorflow. The simulation parameters and EV specifications used are listed in Table~\ref{tablesim1} and Table~\ref{tableEV} respectively, which align with other EV and transport related research and standards \cite{demand2}, \cite{force}, \cite{roadgrade} - \cite{aero}. A 20\,km long highway road consisting of three lanes with three types of EVs including car, bus and lorry following Krauss mobility model \cite{Krauss}, as shown in Fig.~\ref{figSUMO}, is simulated in Simulation of Urban Mobility (SUMO) \cite{SUMO} framework for data collection. The motivation of simulating highway traffic is due to the higher average speed of vehicles at highways as compared to streets, which leads to increased energy consumption \cite{speed}. In this way, the proposed solution of achieving net-zero through prosumerism can be envisioned for situations when the average demand of a transportation network is high. The standard maximum speed limits of vehicles in a UK highway are followed, i.e., 112.65\,km/hr for cars and buses, and 96.56\,km/hr for lorries. $L_i$ and $l_{S_i}$ are randomly generated by lognormal probability distribution with variance of 1 and mean of 10\,km and 6\,km respectively according to the statistical findings in existing literature \cite{log1} - \cite{log2}. Evaluation results are averaged over 100 simulation runs. 

\subsection{ML based Energy Prediction}
Fig.~\ref{figMLa} shows the performance comparison of various ML models analyzed to predict wind, PV energy output and velocity of EVs. As shown in Fig.~\ref{figMLa}\,(a), (b) and (c), the CatBoost model outperforms all other models in terms of training times. The resulting MAE of CatBoost model is at par with the lowest MAE by LGBM for wind and PV energy predication, as shown in Fig.~\ref{figMLa}\,(a) and (b). The RMSE of Catboost model is the lowest among all ML models for velocity prediction of EVs, as shown in Fig.~\ref{figMLa}\,(c). Therefore, considering the optimum performance of CatBoost model and significance of reducing training times for lowering CO$_2$ emissions in computing \cite{computing}, we have used it for the rest of simulations. 

Fig.~\ref{figrenewCont} shows the time-dependent wind and energy output on first day of every month from 08:00 to 16:00 hours. As shown in Fig.~\ref{figrenewCont}, the wind energy output randomly fluctuates throughout the year but is slightly higher in winter months of London. Whereas, the PV energy output is significantly higher in summer than winter, when the dependent parameters such as air temperature and solar irradiance are raised. Also, the PV energy output reaches its peak during each midday when the solar irradiance is expected to be the highest in a day. However, as shown in Table~\ref{tablerenew}, with an average demand of over 7\,MWh and $N_{EV}=500$, wind and PV sources together can fulfill only 1\% of the total demands of a transportation network with $N_{wind} = N_{PV} = 100$.  Therefore, additional solutions such as prosumerism are equally important to completely achieve net-zero balance in a transportation network. Furthermore, the number and locations of wind and PV systems must also be optimally planned in a region to yield the maximum benefits of renewable resources.

\subsection{Utility of EVs}
Fig.~\ref{figpEV} shows the effect of $p_{EV}$ on $U_i$ of EV $i$. It can be seen in Fig.~\ref{figpEV}\,(a) that $U_i$ rises proportionally with $p_{EV}$. A low $p_{EV}$ results in $U_i<0$. According to Proposition 1 of Prisoner's dilemma game, an EV $i$ with $S_i$ will not act selfish only when $U_i \geq 0$. Also, according to Proposition 2, $U_i \geq RC - \frac{RC}{N_Coop}$ is essential as an additional security so that EV does not conspire to play selfish. Here $N_{Coop}$ is the average number of EVs offering $S_i$ over 100 simulation runs. The higher $U_i$ ultimately requires greater $p_{EV}$. However, due to cost minimization objective of Problem 1, a grid does not opt to avail $S_i$ if it has to pay high $p_{EV}$ contributing to increased $C_G$. Therefore, the maximum benefits of prosumerism can only be achieved at an optimum $p_{EV}$, which is shown in Fig.~\ref{figpEV}\,(b). A maximum of 86.7\% and 87.7\% of total available surplus energy from EVs can be used to meet demands of a transportation network when $N_{EV}$ is 500 and 100, respectively.

Fig.~\ref{figchargeUi} shows the impact of $l_{Si}$ on $U_i$ and resulting number of $N_{EV}^{Supp}$. With increasing $l_{Si}$, the EVs have to pay more travel cost which results in a low or negative $U_i$. According to Proposition 2, there is a threshold of $U_i$ for EVs to be motivated for cooperation. A low $U_i$ results in small $N_{EV}^{Supp}$, as shown in Fig.~\ref{figchargeUi}. Consequently, the EVs despite having a surplus amount of energy, do not contribute towards grid load reduction. Therefore, it is important that a charging station is nearby so that an EV has to bear low travel cost. Hence, optimal planning of number and locations of charging stations is significant to increase $N_{EV}^{Supp}$.

\begin{figure*}%
	\centering
	\captionsetup{justification=centering}
	\subfloat[\fontfamily{ptm} \selectfont \small $N_{EV}=500$]{{\includegraphics[width=8.5cm]{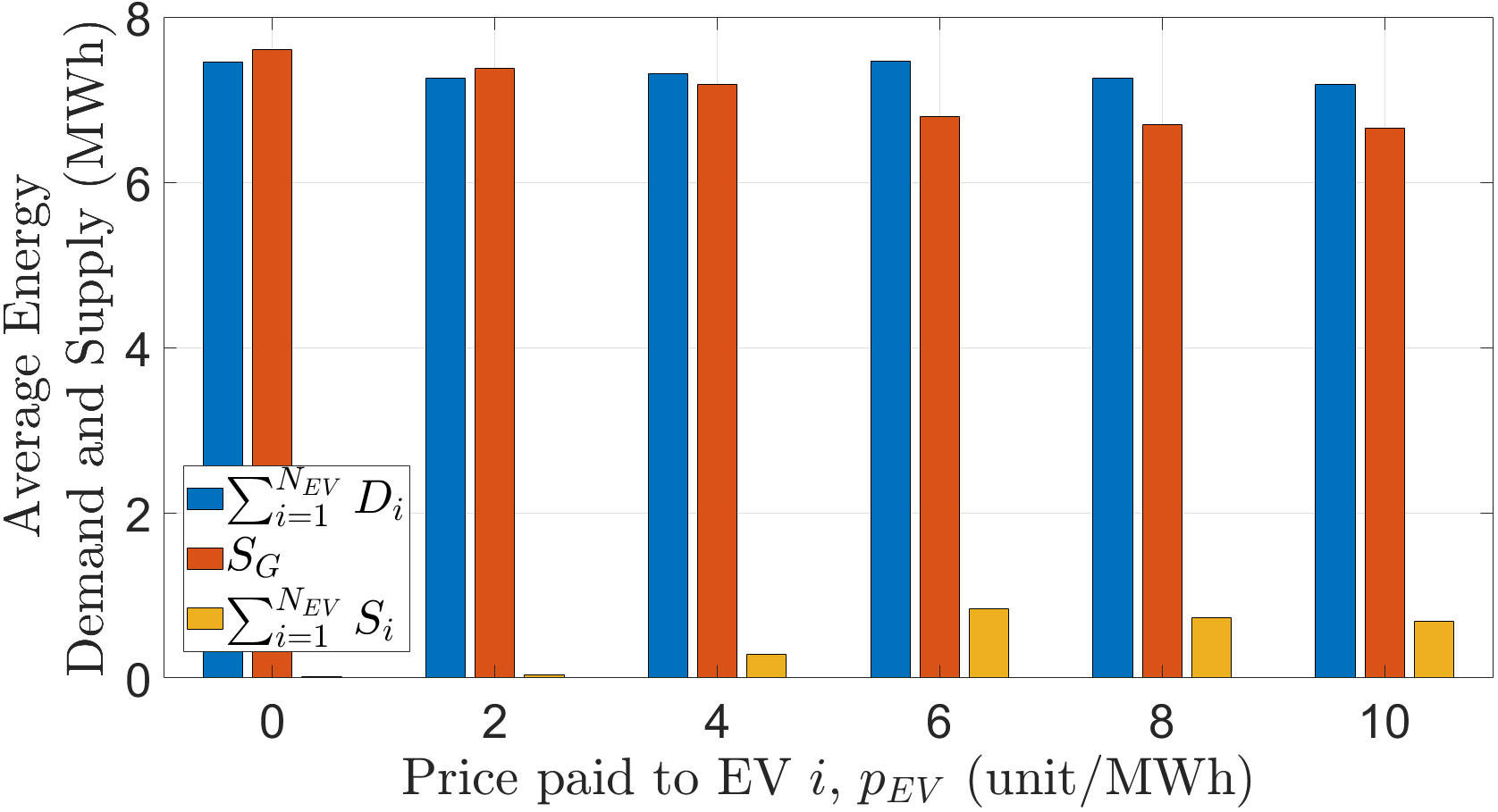} }}%
	\quad
	\subfloat[\fontfamily{ptm} \selectfont \small $N_{EV}=1000$]{{\includegraphics[width=8.5cm]{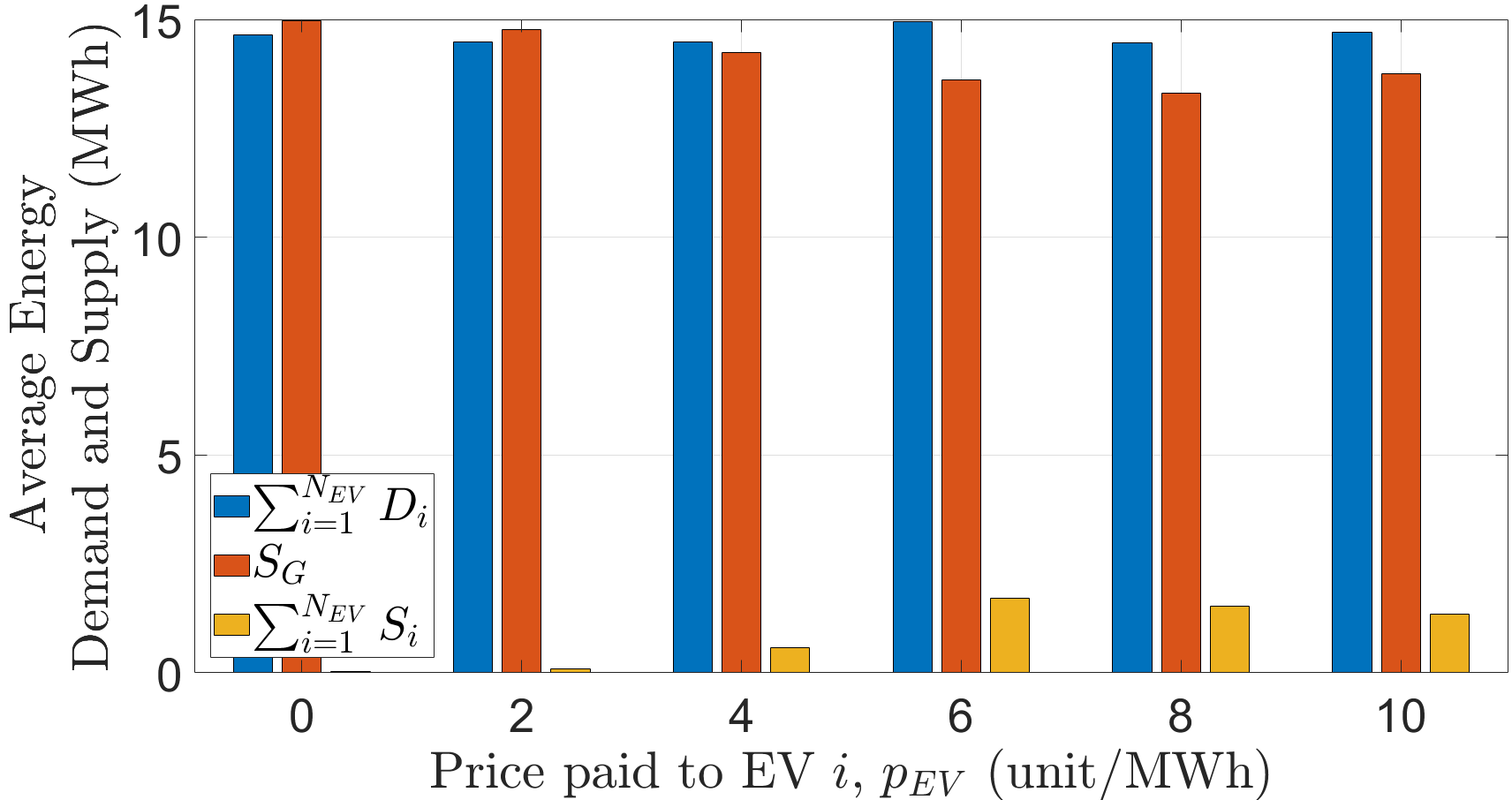} }}%
	\caption{Contribution of prosumerism on optimized supply to transportation network, $m_G=0.05$\,kg, $N_{wind} = N_{PV} = 50$.}
	\label{figEVCont}
\end{figure*}
\begin{figure}%
	\centering
	\captionsetup{justification=centering}
	\includegraphics[width=0.47\textwidth]{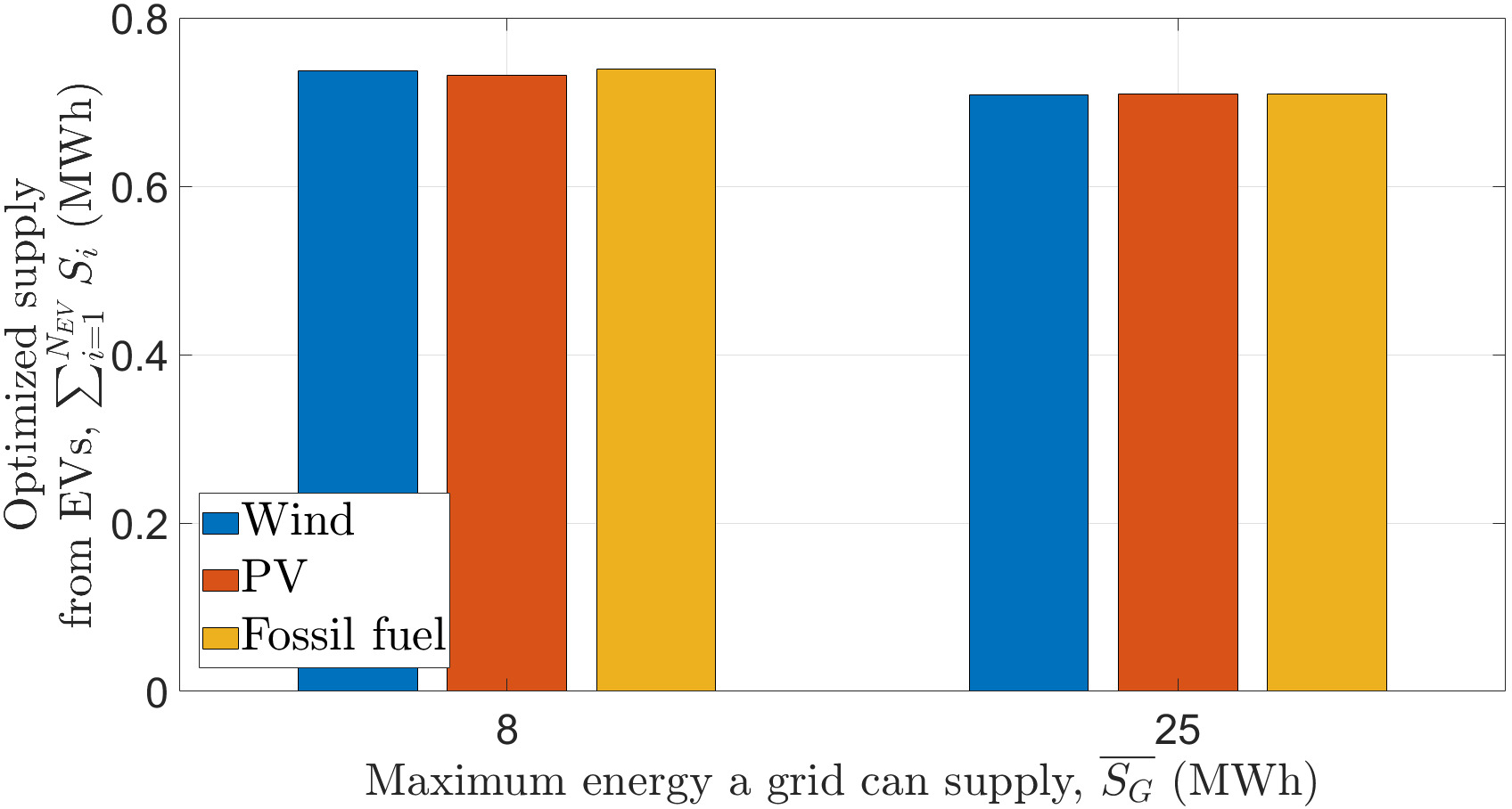}
	\caption{Effect of $\overline{S_G}$ on prosumerism, $N_{EV}=500, N_{wind} = N_{PV} = 50, p_{EV}=6$.}
	\label{figSG}
\end{figure}
\begin{table}[!t]
	\caption{Average Values of $\sum_{w=1}^{N_{wind}}S_w$ and $\sum_{v=1}^{N_{PV}}S_v$. \label{tablerenew}}
	\centering
	\begin{tabular}{|c|c||c|c|}
		\hline
		\textbf{$N_{Wind}$} & \textbf{$\sum_{w=1}^{N_{wind}}S_w$} (kWh) & \textbf{$N_{PV}$} & \textbf{$\sum_{v=1}^{N_{PV}}S_v$} (kWh)  \\
		\hline
		50  & 19.57 & 50 &  18.55 \\
		100  & 38.65 & 100 & 37.10  \\
		\hline
	\end{tabular}
\end{table}
\subsection{Optimal Number of Charging Stations}
Fig.~\ref{figNcharge} shows the number of EVs per charging station as discussed in Section III. Assuming that the EVs and charging stations are uniformly positioned on a road, $N_{EV}^{Dem}$ is inversely proportional to $N_{CS}$. On the contrary, $N_{EV}^{Supp}$ varies according to $l_{S_i}$ that an EV $i$ has to travel to supply $S_i$. For a small $N_{CS}$, $l_{S_i}$ is large which results in low $U_i$ which discourages EVs to become sellers. For large $N_{CS}$, the EVs are distributed proportionally to each charging station. However, an adequate influx of EVs at a charging station is economically beneficial for its optimum utilization and associated installation and maintenance cost. Fig.~\ref{figNcharge} shows the optimal number of charging stations to gain maximum $N_{EV}^{Supp}$.

\subsection{Contribution of EVs as prosumers}
The contribution of EVs as prosumers, i.e., the amount of optimized $\sum_{i=1}^{N_{EV}} S_i$ received from EVs by the grid is dependent upon the source of energy produced by the grid. As shown in (\ref{eqcost}), a grid has to pay $PC$ as a penalty for producing per unit mass of CO$_2$ released during energy generation. Fig.~\ref{figenergysource} shows the impact of energy source on optimized prosumerism. A fossil fuel based energy source produces more CO$_2$ as compared to wind and PV sources \cite{carbon}. Therefore, if a grid provides fossil fuel based $S_G$, it aims to minimize its cost by buying more $S_i$ from an EV $i$ to fulfill the demands of a transportation network than generating its own energy, even when $p_{EV}$ is high. $C_G$ is therefore decreased by reducing high amount of penalty paid to regulation authorities for CO$_2$ emissions. With wind, PV and fossil fuel based energy source, the optimized $\sum_{i=1}^{N_{EV}} S_i$ is 69.63\%, 69.78\% and 99.79\% of the total surplus energy available in EVs respectively. It shows that the proposed solution promotes the usage of environment-friendly energy sources for cost optimization and reduces CO$_2$ emissions through prosumerism, provided that some restrictions are imposed by the regulation authorities. 

Fig.~\ref{figEVCont} shows the amount of contribution made by prosumerism to fulfill the demands of a transportation network when the grid generates energy from a wind source with $m_G=$0.05\,kg. At $p_{EV}=6$, the EVs can fulfill a maximum of 11.29\% and 11.47\% of $\sum_{i=1}^{N_{EV}} D_i$ when $N_{EV}$ is 500 and 100, respectively. With varying $p_{EV}$ and $N_{EV}$, the prosumerism by EVs can contribute on an average of 5.96\% of the total demand requirements, even when the energy source at grid produces least CO$_2$ emissions. Additionally, the impact of prosumerism also depends upon the maximum supply available at the grid, i.e., $\overline{S_G}$, which can be seen in Fig.~\ref{figSG}. On an average, the optimized $\sum_{i=1}^{N_{EV}} S_i$ from EVs is 26.67 MWh higher when $\overline{S_G} = 8$\,MWh as compared to $\overline{S_G}=25$\,MWh. Therefore, it can be concluded that prosumerism can play an effective role when the demands of a transportation network are very high, which cannot be met by the grid alone, for example in situations when the grid produces energy from renewable sources and its supply is limited.
\begin{figure*}%
	\centering
	\captionsetup{justification=centering}
	\subfloat[\fontfamily{ptm} \selectfont \small Demand]{{\includegraphics[width=8.5cm]{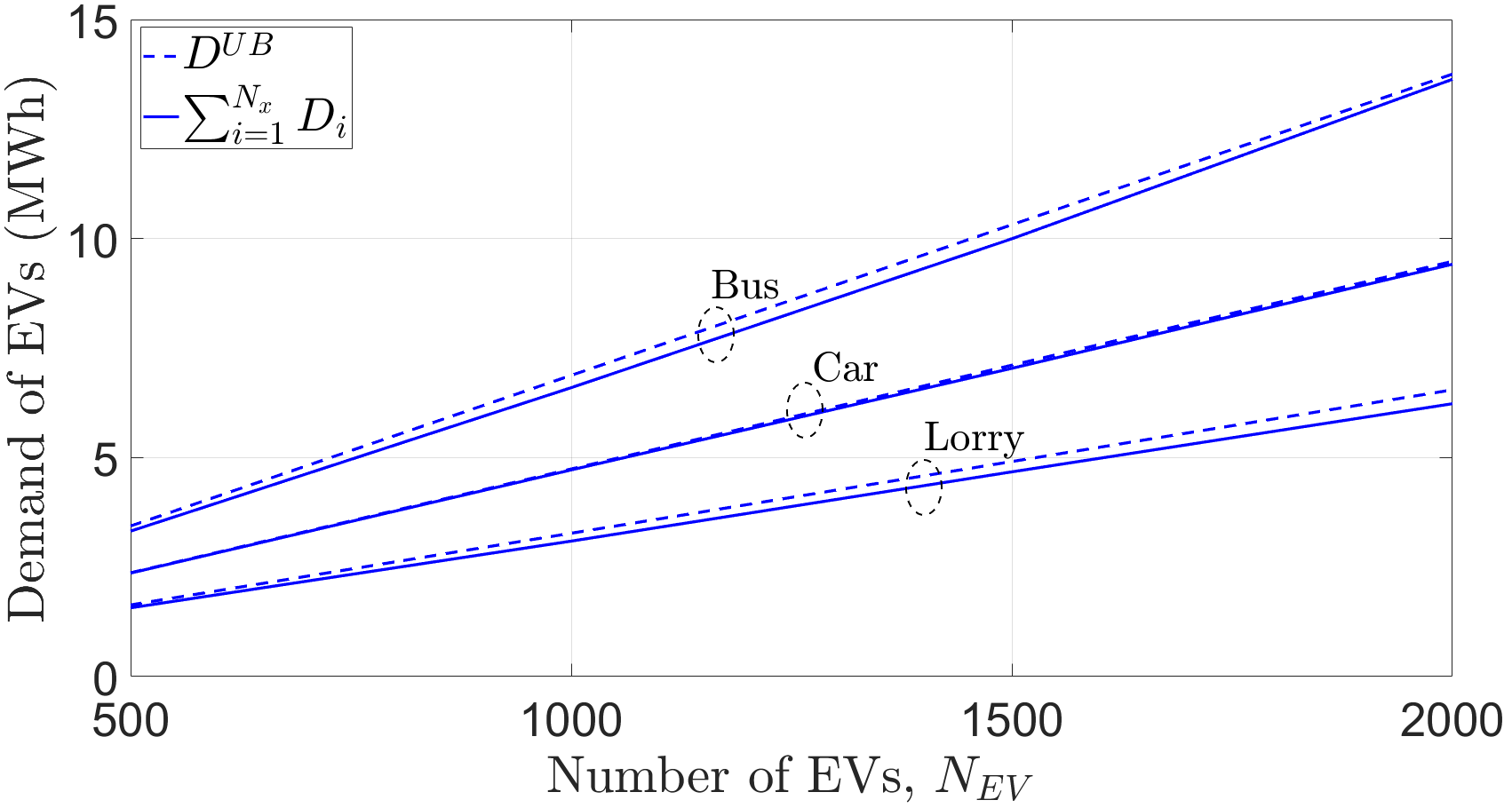} }}%
	\quad
	\subfloat[\fontfamily{ptm} \selectfont \small Supply]{{\includegraphics[width=8.5cm]{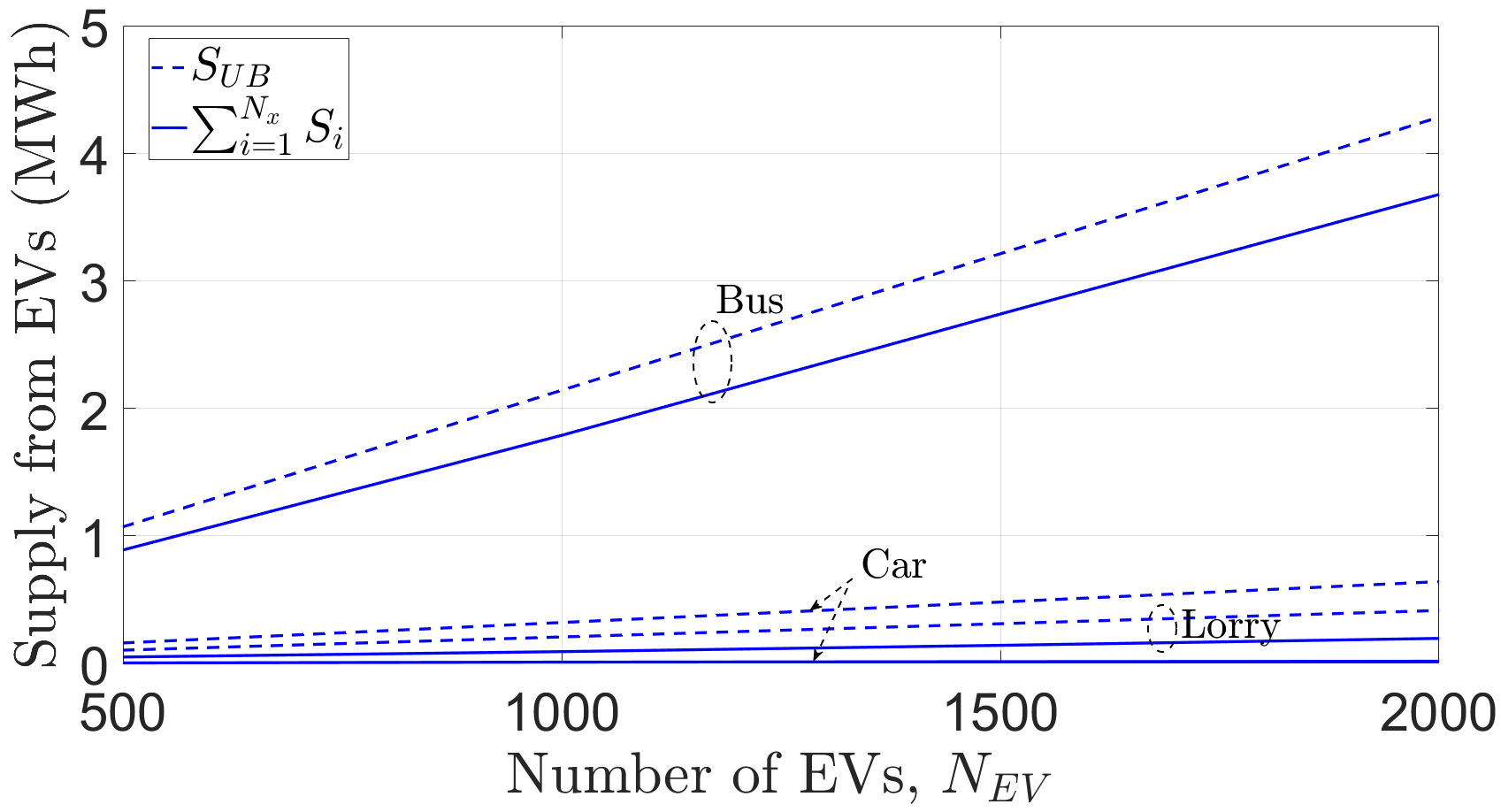} }}%
	\caption{Average demand and supply with respect to EV type.}
	\label{figEVtyp1}
\end{figure*}
\begin{figure*}%
	\centering
	\captionsetup{justification=centering}
	\subfloat[\fontfamily{ptm} \selectfont \small Demanding EVs, $N_{EV}^{Dem}$]{{\includegraphics[width=8.5cm]{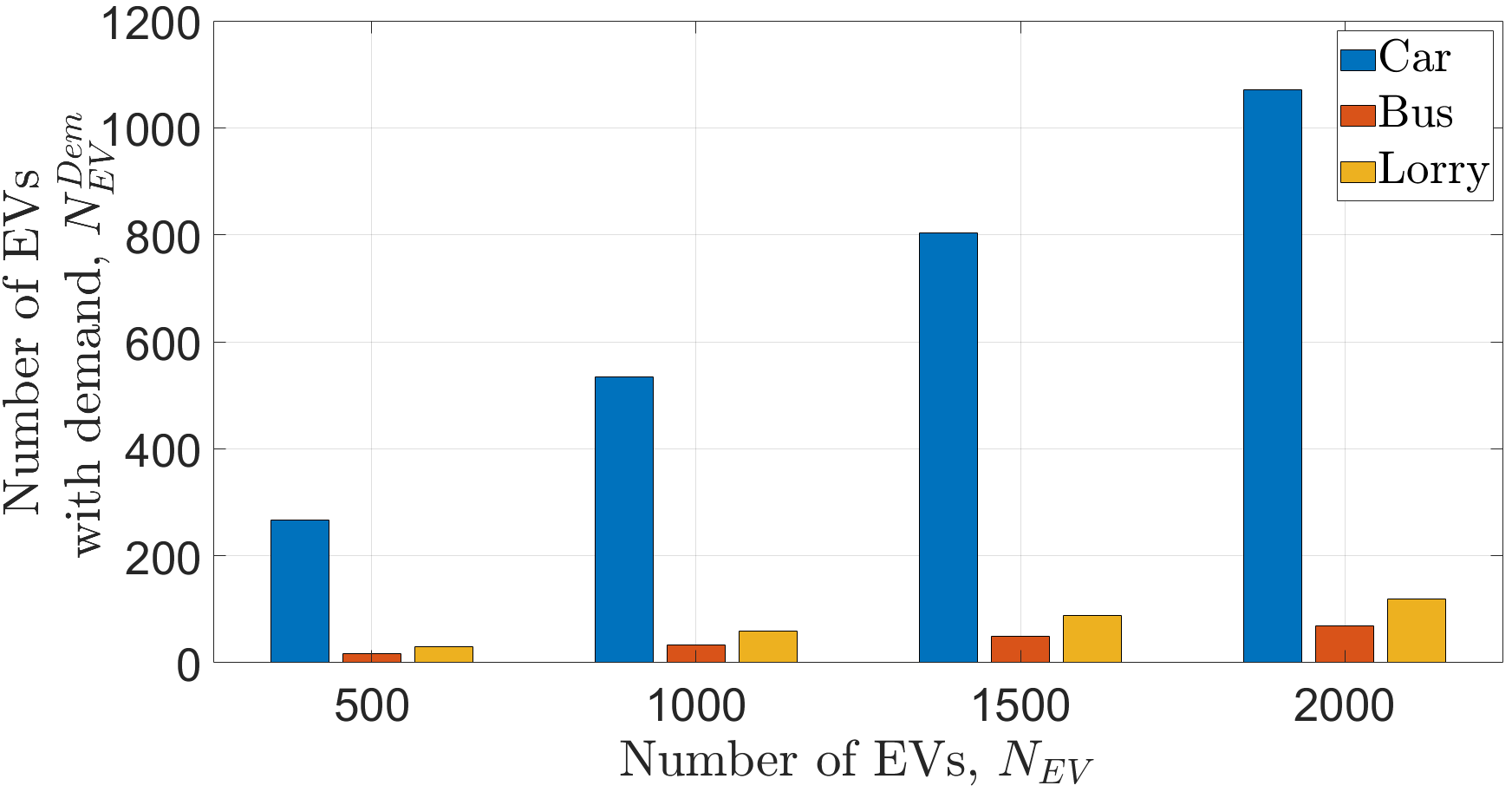} }}%
	\quad
	\subfloat[\fontfamily{ptm} \selectfont \small Supplying EVs, $N_{EV}^{Supp}$]{{\includegraphics[width=8.5cm]{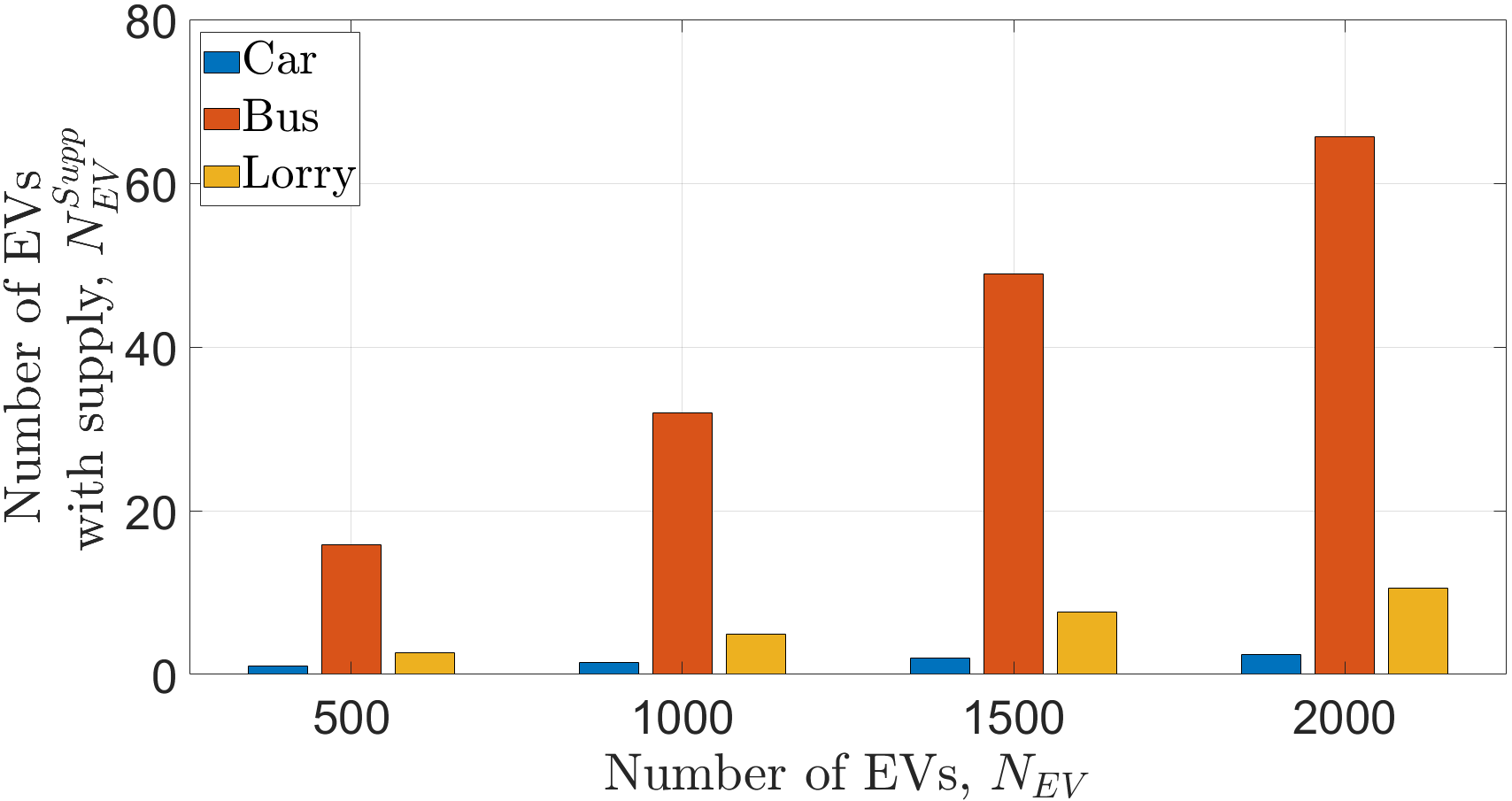} }}%
	\caption{Average number of EVs demanding or supplying energy.}
	\label{figEVtyp2}
\end{figure*}
\begin{figure*}%
	\centering
	\captionsetup{justification=centering}
	\subfloat[\fontfamily{ptm} \selectfont \small Amount of demand $\sum_{i=1}^{N_{EV}} D_i$ and supply $S_G$]{{\includegraphics[width=8.5cm]{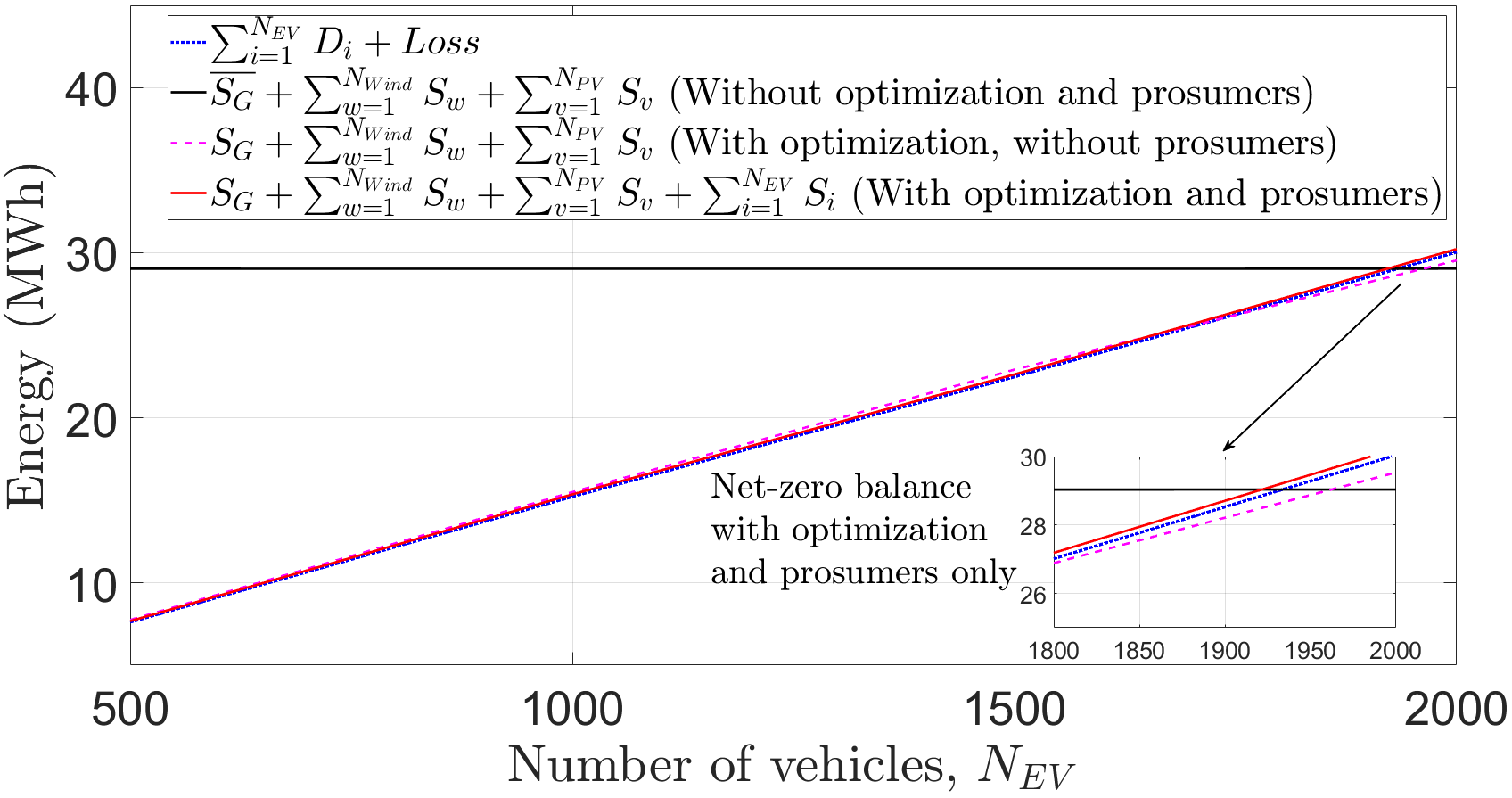} }}%
	\quad
	\subfloat[\fontfamily{ptm} \selectfont \small Grid cost to supply $S_G$]{{\includegraphics[width=8.5cm]{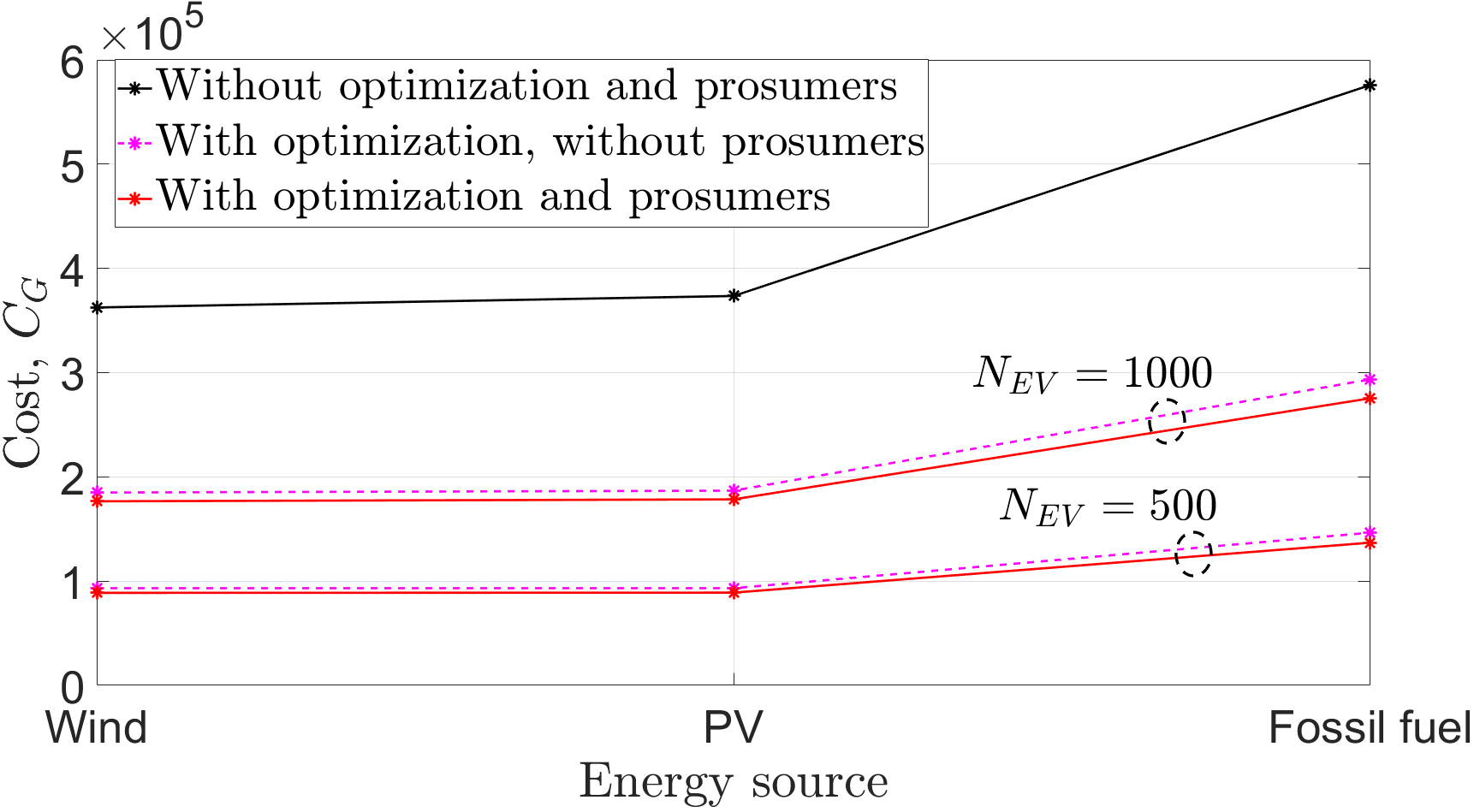} }}%
	\caption{Comparison of the proposed solution without optimization and/or prosumerism, $m_G = 0.05$\,kg, $p_{EV}=6, \overline{S_G}=29$\,MWh.}
	\label{figComp}
\end{figure*}

\subsection{Demand and Supply according to EV type}
Fig.~\ref{figEVtyp1} shows the distribution of demand and supply according to the types of EVs. The theoretical upper bounds $D^{UB}$ and $S^{UB}$ derived in Section III match with the simulations, as shown in Fig.~\ref{figEVtyp1}. $D^{UB}$ and $S^{UB}$ estimations can be used to energy demand and supply if the aggregator loses connectivity with EVs. As shown in Fig.~\ref{figEVtyp1}\,(a) and Fig.~\ref{figEVtyp1}\,(b), an electric bus has the highest demand and supply due to its largest battery capacity, i.e., $BC_i = 320$\,kWh. A lorry has the lowest demand due to its largest $MP_i=220$\,kW. An electric car can contribute least in supplying $S_i$ with smallest $BC_i = 40$\,kWh. 

The number of demanding and supplying EVs are shown in Fig.~\ref{figEVtyp2}. The number of cars with $D_i>0$ is the highest and with $S_i>0$ is the lowest, as shown in Fig.~\ref{figEVtyp2}\,(a) and (b) respectively. On the contrary, the number of buses with $D_i>0$ is least among all EV types and with $S_i>0$ is the highest. It depicts that an electric bus can potentially act as one of the highest supplying prosumers of a transportation network. 

\subsection{Reduction in Grid Load and Cost}
Fig.~\ref{figComp} shows the comparison of the proposed solution without optimization and/or prosumerism. As shown in Fig.~\ref{figComp}\,(a), with no optimization and prosumerism, a grid will provide a constant $\overline{S_G}$ irrespective of $\sum_{i=1}^{N_{EV}} D_i$ in a transportation network, which is a significant waste of resources and cost when $N_{EV}$ is low. Also, in case of high $N_{EV}$, $\overline{S_G}$ without optimization and prosumerism may not be able to achieve demand-supply balance defined in (\ref{netzero}). Failure in demand-supply balance is also possible in optimization without prosumerism when the network demand is high. The supply can be both dynamic and sufficient only when optimization is accompanied with support from EV prosumers. It can be seen in  Fig.~\ref{figComp}\,(a) that when $N_{EV}$ is high, the net-zero balance is only achieved by our proposed solution of optimized prosumerism. Fig.~\ref{figComp}\,(b) compares the impact of the proposed solution on $C_G$, which is highest without optimization and prosumerism. Optimization without prosumerism can reduce $C_G$. However, the proposed optimized solution combined with prosumerism results in a further reduction of averagely 5.3\% in $C_G$. Also, the proposed solution results in 50.97\% lower average cost with wind and PV based energy than fossil fuels. 

\begin{table}[!t]
	\caption{Average Percentage Decrease in Grid Load \label{tablecomp}}
	\centering
	\begin{tabular}{|c|c|}
		\hline
		\textbf{Approach} & \textbf{Reduction in grid load (\%)}  \\
		\hline
		Providing EV's surplus supply  & \multirow{2}*{2.43}  \\
		back to grid \cite{pricing2} & \\
		\hline
		Time-based charging price \cite{pricing2} &   2.44 \\
		\hline
		Price-based charging   &  \multirow{2}*{6.93}  \\
		schedule optimization \cite{Charging2} & \\
		\hline
		Saving EV's surplus supply  &  \multirow{2}*{33.45}  \\
		in battery storage \cite{opt1} & \\
		\hline
		\textbf{Ours} & \textbf{38.21}   \\
		\hline
	\end{tabular}
\end{table}

Table~\ref{tablecomp} compares the results in terms of grid load reduction with other approaches existing in literature. The proposed solution reduces grid load by 38.21\% compared with the load without optimization and prosumerism. Receiving surplus energy from EVs to reduce grid load has also been proposed in \cite{opt1} and \cite{pricing2}. Significant results utilizing battery storage systems are obtained in \cite{opt1}. However, battery storage systems are costly and require maintenance due to battery degradation. The additional advantage of our proposed solution is cost minimization. The saved cost can therefore be utilized to install infrastructure such as V2X network and 5G-enabled aggregator nodes, whose scope is not only limited to energy management but can also be used in various applications of 5G-enabled connected vehicle networks.

\subsection{Communication and Computation Complexity}
Table~\ref{tablecomp2} compares the asymptotic communication and computation complexities of the proposed solution with other approaches utilizing 5G V2X network. A peer to peer communication of EVs in a non-cooperative game setting is presented in \cite{VNC1} and \cite{game1} which results in an increased communication and computation overhead. The demanding and supplying EVs negotiate their energy demands and prices themselves to maximize their utilities in \cite{VNC1}, whereas, in \cite{game1}, each demanding EV sends energy request to $N^{Supp}$ selling aggregators and $N^{Supp}$ selling aggregators share charging slots for $N_{EV}^{Dem}$ EVs, thereby running two communication rounds. For computation, each demanding EV selects the most suitable supplier among $N^{Supp}$ selling aggregators. 

On the contrary, the aggregator based communication and optimization is proposed in \cite{pricing} for V2G energy exchange. The EVs trade their energy with the grid at fixed price and only communicate their demand or surplus energy once before the aggregator runs the algorithm to maximize the utilities of all demanding and supplying EVs. Similarly, in our proposed solution, the aggregator collects demand and surplus energy information from EVs once and the grid runs the optimization algorithm with constraint C2 which considers the positive utility of a supplying EV. Therefore, both the solutions governed by 5G-enabled aggregators result in least communication and computation overhead.

\begin{table*}[!t]
	\caption{Asymptotic Communication and Computation Complexities of Energy Trading Approaches among EVs \label{tablecomp2}}
	\centering
	\begin{tabular}{|c|c|c|}
		\hline
		\textbf{Approach} & \textbf{Communication Complexity} & \textbf{Computation Complexity} \\
		\hline
		Non-cooperative Stackelberg game \cite{VNC1} &  {$\mathcal{O} ((N_{EV}^{Dem} N_{EV}^{Supp})^2)$} &  {$\mathcal{O} (N_{EV}^{Dem} N_{EV}^{Supp})$}   \\ \hline
		Non-cooperative blockchain consensus \cite{game1} &   {$\mathcal{O} (2N_{EV}^{Dem} N^{Supp})$} &   {$\mathcal{O} (N_{EV}^{Dem} N^{Supp})$} \\ \hline
		Auction mechanism \cite{pricing} & $\mathcal{O} (N_{EV}^{Dem} + N_{EV}^{Supp})$ & $\mathcal{O} (N_{EV}^{Dem} + N_{EV}^{Supp})$ \\ \hline
		Ours &  $\mathcal{O} (N_{EV}^{Dem} + N_{EV}^{Supp})$ & $\mathcal{O} (N_{EV}^{Dem} + N_{EV}^{Supp})$ \\
		\hline
	\end{tabular}
\end{table*}

\section{Conclusion}
This paper proposes a smart energy management approach to minimize grid cost and optimize prosumerism. Various ML models are analyzed to predict wind and PV energy output, and EVs' velocity to forecast their demands and surplus supplies. The CatBoost model is found as the optimum solution. Branch and bound based MILP solution is proposed for grid cost minimization and dynamically regulate its supply considering the energy supplied from EVs. Furthermore, the paper has presented theoretical analysis of incentive distribution mechanism using prisoner's dilemma, upper bounds of demands and supply and expected number of demanding and supplying EVs per charging station leading to the optimized number of charging stations on a given length of road. The proposed solution results in an average of 38.21\% reduction in grid supply compared with the supply without optimization and prosumerism. Additionally, the EVs acting as prosumers can reduce 5.3\% of grid cost on average, compared with optimized supply from grid without prosumers. Also, the penalty charge for CO$_2$ emissions proposed in the solution results in less than 50\% of the cost as compared to using fossil fuels, thereby encouraging the use of renewable resources. The communication and computation complexity of the proposed solution is found to be less than the non-cooperative energy trading approaches among EVs. Future research directions include time-based traffic predictions and optimization, investigation of optimised locations and number of wind and PV systems, and exploring detailed usage and challenges of V2G technology associated with sharing supply from prosumers of one region or network to another.



%

\newpage

 




\vfill

\end{document}